\documentclass[final]{IEEEtran}

\usepackage[utf8]{inputenc}

\usepackage{stmaryrd} %paréntesis de corchete dobles
\usepackage{cite} %multiple references at once

\usepackage{amsmath}
\usepackage{amssymb}
\usepackage{amsfonts}
\usepackage{amscd}
\usepackage{mathtools}  %to define floor and ceil bellow

\usepackage{tabularx}
\usepackage{hyperref}

\usepackage{dsfont} % for bold 1

\usepackage{pgf,tikz}  %for figures
\usetikzlibrary{arrows} 
\usepackage{tikz-cd}  %Commutative diagrams

\newcommand{\Exx}[2]{\mathbb{E}_{#1}\left(#2\right)}
\renewcommand{\Pr}[1]{\mathbb{P}\left(#1\right)}

\newcommand{\Ber}{\operatorname{Ber}}
\newcommand{\salg}[1]{\mathfrak {#1}}

\newcommand{\cf}{\mathds{1}}
\newcommand{\cmeas}{\nu^{\#}}

\newcommand{\diff}[2]{\frac{\mathrm{d}  #1}{\mathrm{d}  #2}}

\newcommand{\norm}[1]{\left\|#1\right\|}

\newcommand{\absval}[1]{\left|#1\right|}

\def \d{\mbox{\(\,\mathrm{d}\)}}
 
\renewcommand{\epsilon}{\varepsilon}

\newcommand{\sm}{\smallsetminus}
\newcommand{\Lip}{\operatorname{Lip}}

\newcommand{\Rr}{\mathbb{R}}

\newcommand{\Nn}{\mathbb{N}}

\newcommand{\Zz}{\mathbb{Z}}

\newcommand{\set}[2]{\{\,#1 \, : \, #2\,\} }
\newcommand{\bigset}[2]{\left\{\,#1 \, : \, #2\,\right\} }

\newcommand{\ent}[1]{H_{#1}}
\newcommand{\e}{\mathrm{e}}
\newcommand{\diam}{\operatorname{diam}}
\newcommand{\keyt}[1]{\emph{#1}}
\newcommand{\V}[1]{\mathbf{#1}}

\newcommand{\mres}{\mathbin{\vrule height 1.6ex depth 0pt width
0.13ex\vrule height 0.13ex depth 0pt width 1.3ex}}

\newtheorem{theorem}{Theorem}
\newtheorem{definition}{Definition}
\newtheorem{proposition}{Proposition}
\newtheorem{lemma}{Lemma}
\newtheorem{corollary}{Corollary}
\newtheorem{remark}{Remark}
\newtheorem{example}{Example}

\begin{document}
\title{Typicality for stratified measures}
\author{Juan Pablo Vigneaux
\thanks{Juan Pablo Vigneaux is with the Department of Mathematics, California Institute of Technology, Pasadena, CA 91125, USA (e-mail:vigneaux@caltech.edu).}}

%\thanks{Manuscript received xxx; revised xxx; accepted 

\maketitle

\begin{abstract}
\emph{Stratified measures} on Euclidean space are defined here as convex combinations of rectifiable measures. They are possibly singular  with respect to the Lebesgue measure and generalize continuous-discrete mixtures. A stratified measure $\rho$ can thus be represented as $\sum_{i=1}^k q_i \rho_i$, where $(q_1,..,q_k)$ is a probability vector and each $\rho_i$ is $m_i$-rectifiable for some integer $m_i$ i.e. absolutely continuous with respect to the $m_i$-Hausdorff measure $\mu_i$ on a $m_i$-rectifiable set $E_i$ (e.g. a smooth $m_i$-manifold).
We introduce a set of strongly typical realizations of $\rho^{\otimes n}$ (memoryless source) that occur with high probability. The typical realizations are supported on a finite union of strata $\{E_{i_1}\times \cdots \times E_{i_n}\}$ whose dimension concentrates around the \emph{mean dimension} $\sum_{i=1}^k q_i m_i$. For each $n$, an appropriate sum of Hausdorff measures on the different strata gives a natural notion of reference ``volume''; the exponential growth rate of the typical set's volume is quantified by Csiszar's \emph{generalized entropy} of $\rho$  with respect to $\mu=\sum_{i=1}^k \mu_i$. Moreover, we prove that this generalized entropy satisfies a chain rule and that the conditional term is related to the volume growth of the typical realizations in each stratum.  The chain rule and its asymptotic interpretation hold in the more general framework of \emph{piecewise continuous measures}: convex combinations of measures restricted to pairwise disjoint sets equipped with reference $\sigma$-finite measures. Finally, we establish that our notion of mean dimension coincides with R\'enyi's information dimension when applied to stratified measures, but the generalized entropy used here differs from R\'enyi's dimensional entropy. 
\end{abstract}

\begin{IEEEkeywords}
generalized entropy, asymptotic equipartition property, stratified measures, rectifiable measures, rectifiable sets, information dimension, Hausdorff dimension, chain rule, measure concentration
\end{IEEEkeywords}

%\tableofcontents

%\begin{center}
%{\it This paper has been submitted to the IEEE Transactions on Information Theory.}
%\end{center}

\section{Introduction}\label{sec:intr}

Let $(E,\salg E, \mu)$ be a  $\sigma$-finite measure space and $\rho$ a probability measure defined on it. We suppose that $\rho$   is absolutely continuous with respect to $\mu$ (denoted $\rho\ll \mu$), which  is equivalent to the existence of an integrable function $f:E \to \Rr $ such that $\rho(A)= \int_A f\d \mu$ for every $A\in \salg E$ \cite[Thm. 4.2.2.]{Cohn2013}. Any such $f$ is said to be a version of the Radon-Nikodym derivative $\d\rho/\mathrm{d}\mu$; let us fix one of these versions. In this setting, the generalized entropy of $\rho$ with respect to $\mu$ is
\begin{equation}\label{eq:generalized_entropy}
\ent \mu(\rho) := -\int_{E}\ln \diff{\rho}{\mu}\d \rho = \Exx{\rho}{-\ln \diff{\rho}{\mu}}. 
\end{equation}
 This entropy was introduced by Csisz\'ar  \cite{Csiszar1973} in connection with the problem of quantization of signals. 
 From a probabilistic perspective, it is also  justified by the asymptotic equipartition property (AEP): for any $\epsilon > 0$ and $\delta>0$, the set of weakly $\delta$-typical realizations
\begin{multline}
W^{(n)}_\delta(\rho) = \\\bigset{(x_1,...,x_n)\in E^n}{\absval{-\frac 1n\ln \prod_{i=1}^n \diff\rho\mu(x_i) - \ent \mu(\rho)} < \delta}
\end{multline}
has $\rho^{\otimes n}$-measure greater than $1-\epsilon$,  provided $n$ is big enough, and its ``size'' or ``volume'' $\mu^{\otimes}(W^{(n)}_\delta(\rho))$ satisfies then the inequalities
\begin{equation}\label{eq:asymptotic_volume}
(1-\epsilon) \e^{n(\ent \mu(\rho) -\delta)} \leq \mu^{\otimes}(W^{(n)}_\delta(\rho))\leq \e^{n(\ent \mu(\rho) +\delta)}.
\end{equation}
To simplify the presentation, we occasionally write  $\mu^{\otimes n}(W_{\delta}^{(n)}(\rho)) \approx \exp(n H_{\mu}(\rho))$ instead of \eqref{eq:asymptotic_volume}.

The discrete entropy and differential entropy introduced by Shannon \cite{Shannon1948} arise as particular cases of \eqref{eq:generalized_entropy}: the former when $\mu$ is the counting measure on a discrete set equipped with the $\sigma$-algebra of all subsets; the latter when $E=\Rr^d$, $\salg E$ is generated by the open sets, and $\mu$ is the Lebesgue measure. Up to a sign, the Kullback-Leibler divergence---also called \emph{relative entropy}---is another particular case, when $\mu$ is a probability law on an arbitrary measure space.

Another remarkable case was treated recently by   Koliander, Pichler, Riegler, and Hlawatsch  \cite{Koliander2016}: the entropy of an $m$-rectifiable probability measure with respect to the $m$-Hausdorff measure. A Radon measure $\rho$ on $\mathbb R^n$ is called $m$-rectifiable if the ``density''
\begin{equation}
\Theta_m(\rho,x) = \lim_{r\downarrow 0} \frac{\rho(B(x,r))}{w_mr^m}
\end{equation}
exists and is finite and nonzero for $\rho$-almost every $x$; here $B(x,r)$ denotes the Euclidean ball of radius $r$ centered at $x$ and $w_m$ is the Lebesgue measure of the unit ball $B(0,1)$ in $\Rr^m$. In virtue of celebrated theorem by Preiss \cite{DeLellis2008}, the preceding condition is equivalent to the existence of an $m$-rectifiable set $E$---essentially, a subset of a countable union of $m$-dimensional $C^1$-submanifolds of $\Rr^d$---such that $\rho \ll \mathcal H^m \mres E$. In this statement, $\mathcal H^m$ is $m$-Hausdorff measure, the most natural notion of $m$-dimensional volume in Euclidean space, and $\mathcal H^m\mres E$ denotes its restriction to $E$, which is $\sigma$-finite. Moreover, $\Theta_m(\rho,\cdot)$ is a version of the Radon-Nikodym derivative $\d\mu/\d( \mathcal H^m \mres E)$ \cite[Thm. 2.83]{Ambrosio2000}.

In this article, we study mixtures (convex combinations) of rectifiable measures on $\Rr^d$ from the probabilistic viewpoint of the AEP. Although this turned out to be the most natural setting for our results, initially we were simply interested in discrete-continuous mixtures. The latter are measures $\rho$ that satisfy
\begin{equation}\label{eq:simple_mixture}
\rho = q \rho_{ac} + (1-q) \rho_{pd},
\end{equation}
where $q\in [0,1]$, $\rho_{ac}$ is a probability measure \emph{absolutely continuous} with respect to the Lebesgue measure $\lambda^{d}$, and $\rho_{pd}$ is a \emph{purely discontinuous} probability measure i.e. of the form $\sum_{e\in E_0} p_e \delta_{e}$ for some countable subset $E_0$ of $\Rr^d$. In that case, $\rho_{pd}$ is absolutely continuous with respect to (w.r.t.) the counting measure $\mu_0$ on $E_0$, whereas $\rho \ll \mu:=\lambda + \mu_0$, so it makes sense to talk about $H_{\mu}(\rho)$. Moreover, the following ``chain rule'' holds (Lemma \ref{lem:chain_rule}):
\begin{equation}\label{eq:entropy_simple_misture}
\ent \mu (\rho) = \ent{\#}(q,1-q) + q \ent\lambda (\rho_{a.c.}) + (1-q)  \ent{\mu_0}(\rho_d). 
\end{equation}
We denote by $\ent\#$ the discrete Shannon entropy when the reference counting measure is clear from context.

 R\'enyi \cite{Renyi1959} \emph{defined} the entropy of the mixture \eqref{eq:simple_mixture} as the right-hand side of \eqref{eq:entropy_simple_misture} following a very different reasoning. Given an arbitrary probability measure $\rho$ on $\Rr^d$, he first discretized it through a measurable partition of $\Rr^d$ into cubes with vertices in $\Zz^d/n$, getting laws $\rho_n$ with countable support; in general, the discrete entropies $\ent{\#}(\rho_n)$  diverge when $n\to \infty$. R\'enyi defines the \emph{information  dimension} of $\rho$ as $s=\lim_n \ent{\#}(\rho_n)/\ln n$, and its \emph{$s$-dimensional entropy} as $\ent{s}^R(\rho) := \lim_n (\ent{\#}(\rho_n) -s \ln n)$, provided these limits exist. In particular, he proved that the dimension of \eqref{eq:simple_mixture} is $qd$ and that $\ent{qd}^R(\rho)$ equals the expression on the right-hand side of \eqref{eq:entropy_simple_misture}, see \cite[Thm.~3]{Renyi1959}. In fact, he considered the particular case $d=1$, but his arguments can be easily adapted to general $d$. The cases $q=1$ and $q=0$ correspond, respectively, to Shannon's differential entropy of a variable of information dimension $d$ and to Shannon's discrete entropy of a variable of information dimension 0.
 
A discretization such as R\'enyi's represents an additional arbitrary datum, not determined by $\rho$. And although R\'enyi  claimed (without proof) that the information dimension of a probability law supported on a smooth submanifold equals the topological dimension of the latter, he did not give a proper geometrical meaning to the information dimension of the simple mixture \eqref{eq:simple_mixture} e.g. as the dimension of a manifold or a fractal set.  In fact, he wondered about the ``topological background'' of a distribution with noninteger dimension \cite[p.~209]{Renyi1959}. It is therefore natural to ask if there is a more intrinsic way of giving a meaning to the formula \eqref{eq:entropy_simple_misture}, directly in terms of the asymptotic behavior of $\rho^{\otimes n}$.

Remark that $\rho^{\otimes n}$ is a probability measure on the $nd$-dimensional vector space $(\Rr^{d})^{\otimes n}$ and that
\begin{equation}\label{eq:sum_decomposition_product_simple_mixture}
\rho^{\otimes n} = \sum_{\V y \in \{0,1\}^n} q^{N(1;\V y)} (1-q)^{n-N(1;\V y)} \rho_{y_1}\otimes \cdots \otimes \rho_{y_n},
\end{equation}
where $N(1;\V y)$ is the number of ones in $\V y = (y_1,...,y_n)\in \{0,1\}^n$, $\rho_1 = \rho_{ac}$, and $\rho_0 = \rho_{pd}$. The term in this sum  indexed by $\V y$ is a probability measure supported on a product $(\Rr^d)^{N(1;\V y)}\times E_0^{n-N(1;\V y)}$ (up to a permutation of the factors) which we call a \emph{stratum}, weighted by $q^{N(1;\V y)}(1-q)^{n-N(1;\V y)}$. We may  consider the vector $\V y$ as a realization of a sequence $\V Y = (Y_1,...,Y_n)$ of i.i.d. random variables with law $\Ber(q)$; the asymptotic equipartition property gives, for any $\epsilon>0$ and $\delta'>0$,  a set $W_{\delta'}^{(n)}(q)$ of typical realizations of $\V Y$ such that, for $n$ sufficiently big, $W_{\delta'}^{(n)}(q) \approx \e^{n H_{\#}(q,1-q)}$ and $\sum_{\V y \in W_{\delta'}^{(n)}(q)} q^{N(1;\V y)}(1-q)^{n-N(1;\V y)} \geq 1-\epsilon$. The typical realizations $\V y \in W_{\delta'}^{(n)}(q)$ have roughly $nq$ ones. The measure
\begin{equation}\label{eq:n-approximation-simple-mixture}
\rho^{(n)} =  \sum_{\V y \in W_{\delta'}^{(n)}(q)} q^{N(1;\V y)} (1-q)^{n-N(1;\V y)} \rho_{y_1}\otimes \cdots \otimes \rho_{y_n},
\end{equation}
is thus a good approximation of $\rho^{\otimes n}$ and the associated strata have roughly dimension $nqd$. This is a geometrical sense in which the dimension of \eqref{eq:simple_mixture} is $qd$: each factor $\rho$ of $\rho^{\otimes n}$ contributes, on average, $qd$ to the ``typical dimension'' $nqd$.

Let us write \eqref{eq:entropy_simple_misture} in multiplicative form,
\begin{equation}\label{eq:entropy_simple_mixture_multiplicative}
\e^{n \ent{\mu}(\rho)} = \e^{n H_\#(q,1-q)} \e^{n(q \ent{\mu_1} (\rho_{1}) + (1-q)  \ent{\mu_0}(\rho_0))},
\end{equation}
using the notation $\mu_1 = \lambda^{ d}$.
We recognize on the left-hand-side an approximation of the $\mu^{\otimes n}$-volume of a set $W_\delta^{(n)}(\rho)$ of typical realizations of $\rho^{\otimes n}$; on the right hand side, $\e^{n \ent\#(q,1-q)}$ approximates the number of typical realizations of $\V Y$, hence of the number of typical strata, so that $\e^{n(q \ent{\mu_1} (\rho_{a.c.}) + (1-q)  \ent{\mu_0}(\rho_d))}$ should approximate the $\mu_{y_1}\otimes \cdots \otimes \mu_{y_n}$-volume of the typical realizations of $\rho^{\otimes n}$ in the stratum associated to a typical $\V y= (y_1,...,y_n)$.   We prove that this is indeed the case in \ref{sec:cond_entrop_asymp}.

 It is convenient to work with a partition $\{E_0,E_1\}$ of the ambient space $\Rr^d$, i.e. $E_1=\Rr^d \setminus E_0$, so that $Y$ becomes a deterministic function of an $\Rr^d$-valued variable $X$ with law $\rho$ via the projection $\pi:\Rr^d\to \{0,1\}$ that maps $E_i$ to $i$.  In this case the stratum $\Sigma_{\V y}$ associated with $\V y\in \{0,1\}^n$ is simply $(\pi^{\times n})^{-1}(\V y)$. We can safely take $\rho_1 = \rho_{ac}\mres E_1$ instead of $\rho_1=\rho_{ac}$ in \eqref{eq:sum_decomposition_product_simple_mixture}, because $E_0$ is $\rho_{ac}$-null; the arguments in the previous two paragraphs also apply to this setting. 
 
We develop  analogous results for \emph{stratified measures}, which are defined in Section \ref{sec:stratified_def} as mixtures of rectifiable measures. Equivalently, we call a measure  $\rho$ on $\Rr^d$ stratified if there is a sequence $ (E_i)_{i=1}^k$ of \emph{disjoint} rectifiable sets of $\Rr^d$ such that $E_i$ is $m_i$-rectifiable, $0 \leq m_1 < m_2 <\cdots <m_k\leq d$, and 
 \begin{equation}\label{eq:stratified_measure}
 \rho = \sum_{i=1}^k q_i \rho_i
 \end{equation}
 for certain probability vector $(q_1,...,q_k)$ and probability measures $\rho_i$ absolutely continuous with respect to the $m_i$-dimensional Hausdorff measure  restricted to $E_i$, which we denote $\mu_i$. In this case, the strata of $\rho^{\otimes n}$ are the products $\Sigma_{\V y} :=E_{y_1}\times \cdots \times E_{y_n}$, for any sequence $\V y = (y_1,...,y_n)\in \{1,...,k\}^n$, and  $\rho^{\otimes n}$ can be written as a weighted sum of probability measures on those strata; the weight in front of the measure associated to $\V y$ is $q_1^{N(1;\V y) }\cdots q_k^{N(k; \V y)}$, where $N(i;\V y)$ is the number of appearances of the symbol $i$ in $\V y$. 
 
One can see $\V y$ as the realization of an i.i.d. sequence $\V Y =(Y_1,...,Y_n)$ of $\{1,...,k\}$-valued random variables such that $P(Y_1 = i) = q_i$ for each $i\in E_Y:=\{1,...,k\}$; the dimension of a stratum is then a random variable $D_n = \sum_{i=1}^n m_{Y_i}$ with mean $\mathbb E(D_n) = n \mathbb E(D_1)= n \sum_{j=1}^k q_j m_j$. The theory of strong typical sequences developed by Csisz\'ar and K\"orner  \cite{Csiszar1981} shows that, for $n$ big enough, there is a set $A_{\delta}^n$ of ``strongly'' typical realizations of $\V Y_n$ such that the dimension of a typical stratum corresponding to any of these typical realizations---this is, $\sum_{i=1}^n m_{y_i}$ or equivalently $\sum_{j=i}^k N(j; \V Y_n) m_j$---is contained in an interval 
 $$[n \mathbb E(D_1) - O(n^{1/2+\xi}), n\mathbb E(D_1) + O(n^{1/2+\xi})],$$
for any $\xi>0$, provided $n$ is large enough.  In other words, for any $\V y\in A_\delta^n$, the dimension of the associated stratum $\Sigma_{\V y}$ satisfies $\dim \Sigma_{\V y}/n = \mathbb E(D_1) + O(n^{-1/2+\xi})$. In view of this, it makes sense to define the dimension of the stratified measure  \eqref{eq:stratified_measure} as $\mathbb E(D_1)=\sum_{j=1}^k q_j m_j$; this is the asymptotic contribution by each factor $\rho$ in $\rho^{\otimes n}$ to the average dimension of a typical stratum.
 
The stratified measure $\rho$ is absolutely continuous with respect to $\mu'=\sum_{i=1}^k \mu_k$, and its entropy is
\begin{equation}
\ent{\mu'}(\rho) = \ent\#(q_1,...,q_k) + \sum_{i=1}^k q_i  \ent{\mu_i}(\rho_i).
\end{equation}
Again, the $\ent\#(q_1,...,q_k)$ is the exponential rate of growth of typical realizations of the variable $Y$ with categorical distribution $(q_1,...,q_k)$ and $\exp(n(\sum_{i=1}^k q_i  \ent{\mu_i}(\rho_i)))$ approximates the $d(\V y)$-dimensional volume (i.e. the $d(\V y)$-Hausdorff measure) of the typical realizations of $\rho^{\otimes n}$ in the stratum $\Sigma_{\V y}$, for any $\V y\in A_\delta^n$. This is the key contribution of the article, an asymptotic equipartition theorem for stratified measures, precisely stated in Section \ref{sec:typical_dimension}.

 Does our notion of dimension for stratified measures coincide with R\'enyi's information dimension? Although R\'enyi claimed that this is the case when $k=1$ and $E_1$ is a smooth manifold such that $0 <\dim E_1  < d$, he did not provide a proof. 
We shall see that, under a suitable condition, an $m$-rectifiable measure $\rho$ has information dimension $m$; we say that $\rho$ is \emph{dimensional regular}.  The information dimension of the measure  $\rho$ in \eqref{eq:stratified_measure} is $\sum_{i=1}^k q_im_i$, provided each $\rho_i$ is dimensional regular. But, in general, our entropy cannot be recovered as $\lim_n (\ent\#(\rho_n) - d\ln n)$; in fact, there is no reason to expect such a coincidence, since the quantization introduced by R\'enyi is an additional datum and it is not subject to any compatibility condition with the sets $E_i$. 

When we were finishing this manuscript, we found a recent article by Charusaie, Amini, and Rini \cite{Charusaie2022} that studies a special kind of stratified measure that they call \emph{affinely singular}; in this case, each $E_i$ is, up to a set of $m_i$-Hausdorff measure zero, a union of $m_i$-dimensional affine subspaces $\Rr^d$. They even consider briefly general mixtures of rectifiable measures in Section V.A. Their main contribution is a novel information measure called \emph{dimensional-rate bias}, related to the asymptotic behavior of the rate-distortion function in the high-resolution regime $D\to 0$ (the distortion is measured in terms of the squared Euclidean distance). In the case of affinely singular measures, this bias is an entropy in the sense of \eqref{eq:generalized_entropy} \cite[Rem. 3]{Charusaie2022} and it simultaneously generalizes Shannon's discrete and differential entropy. It would be interesting to establish a connection with typicality and to extend their results to  stratified measures using some of the tools developed here.

 \subsection*{Organization of the paper}
 
 Section \ref{sec:Preliminaries} covers some preliminary material that plays a key role in the article: first, a general statement of the asymptotic equipartition property as quoted above; second, the necessary background in geometric measure theory concerning the Hausdorff measure and dimension as well as rectifiable sets. 
 
 Section  \ref{sec:piecewise_cont_defi} introduces piecewise continuous measures, which are certain mixtures of absolutely continuous measures. In Section \ref{sec:chain_rule} we prove that the chain rule holds for their entropy. Section \ref{sec:cond_entrop_asymp} gives an asymptotic interpretation for the conditional term in the chain rule. Section \ref{sec:strong_typicality} summarizes some  aspects of Csisz\'ar and K\"orner's theory of strong typicality; these ideas are used in Section \ref{sec:symmetry} to prove an equipartition result and also in \ref{sec:typical_dimension} to bound the typical dimensions.
 
 Section \ref{sec:rectifiable_measures} reviews some relevant facts about rectifiable measures. Section \ref{sec:stratified_def} contains the definition of stratified measures; they are in particular piecewise continuous. Section \ref{sec:products_rectifiable_measures} explains that products of rectifiable measures are also rectifiable, and Section \ref{sec:typical_strata} establishes the existence of typical strata. Section \ref{sec:typical_dimension} characterizes the dimension of typical strata and states an AEP for stratified measures. 
 
 Finally, Section \ref{sec:Renyi-dim-and-ent}
 discusses the relationship between the dimension and entropy used in this article and those introduced by R\'enyi in \cite{Renyi1959}.
 
 \subsection*{Notation} 
 
 In general, uppercase Roman letters denote sets, but $H$ has been reserved the entropy, $P$ and $Q$ for  probability mass functions, and $X$ and $Y$ for random variables. Lowercase Roman letters denote elements of those sets (for instance, $s\in S$). All measures are denoted by Greek letters, with the only exception of the Hausdorff measure $\mathcal H^m$. Also,  $\varphi$ is always a map; $\pi$ a projection; $\epsilon$ and $\delta$ arbitrarily small positive numbers.
 
 We only use boldface to denote elements of the $n$-fold product of a given set; for instance, $\V x\in E^n$, where $\V x=(x_1,...,x_n)$ and $x_i\in E$. Caveat: in some cases the set $E$ is itself made of vectors (as a subset of $\Rr^d$), but the components of such vectors play no role and the elements of $E$ are written  in regular lowercase. The only exception to this is Section \ref{sec:Renyi-dim-and-ent} at the end of this article, but there we do not consider products $(\Rr^d)^n$. 
 
  Given a subset $A$ of a set $X$, $A^c$ denotes the complement $X\setminus A$; in most cases $X$ is implicit. The inclusion $\subset$ is not necessarily strict.When $X$ is a set, the symbol $2^X$ denoted the set of all subsets of $X$; sometimes we refer to a set of sets as a collection or a family of sets. The characteristic function of the set $A$ is $\cf_A$.The symbols $\Nn$, $\Nn^*$ and $\Rr$ denote respectively the sets of natural numbers $\{0,1,2,...\}$,  positive natural numbers, and  real numbers. 
 
 Integrals are always understood in the Lebesgue sense. We use the standard notations of measure theory; see for instance \cite{Williams1991}. We use uppercase fraktur characters $\salg B, \,\salg E$,... to denote $\sigma$-algebras. If $\mu$ is a measure on $(E,\salg E)$, $\salg B^{\otimes n}$ denotes the $n$-fold product $\sigma$-algebra and  $\mu^{\otimes n}$ denotes the $n$-fold product measure on $(E^{n},\salg B^{\otimes n})$. In turn, $L^1(E,\mu)$ denotes the quotient of the space of $\mu$-integrable functions under the equivalence relation of almost sure equality.
 
 For any $d\in \Nn^*$, we endow the real vector space $\Rr^d$ with the Euclidean norm $\norm{(x_1,...,x_d)}_2 = \sqrt{x_1^2 + \cdots x_d^2}$ and we refer to it as Euclidean space.
 
Other notations are introduced and clarified in the relevant sections.

\section{Preliminaries}\label{sec:Preliminaries}

\subsection{Measures and integration}\label{sec:measure_theory}

We collect here some basic definitions and facts from measure theory that will appear throughout the article. This is unavoidably very condensed; the reader can find detailed explanations in \cite{Cohn2013}, \cite{Williams1991}, and \cite[Ch. I]{Ambrosio2000}. Since the notation is standard, anyone already familiar with the theory may safely skip this section.

A \emph{measurable space} $(E,\salg E)$ is a set $E$ equipped with a $\sigma$-algebra $\salg E$, that is, a collection $\salg E$ of subsets of $E$ such that 
\begin{enumerate}
\item $\emptyset\in \salg E$,
\item if $A$ belongs to $\salg E$, then $A^c = E\setminus A$ belongs to $\salg E$ too;
\item if $(A_1)_{i=0}^\infty\subset \salg E$, then $\bigcup_{i=0}^n A_i$ belongs to $\salg E$.
\end{enumerate}
In probability theory, the set $E$ represents possible outcomes of an experiment, and the sets $A\in \salg E$ are events. 

A \emph{measure} $\nu$ on $(E,\salg E)$ is a function $\nu:\salg E\to [0,\infty]$ such that $\nu(\emptyset)=0$ and $\nu(\bigcup_{i=0}^{\infty} A_i)=\sum_{i=0}^\infty \nu(A_i)$ for any countable collection of pairwise disjoint sets $(A_i)_{i=0}^\infty \subset \salg E$ (a property called \emph{$\sigma$-additivity}); the triple $(E,\salg E,\nu)$ is then called a \emph{measure space}. 

(When the underlying set $E$ is infinite, it is in general not possible to define coherently an additive set function on all its subsets. That explains the appearance of the $\sigma$-algebras, as collections on which one can unambiguously define a measure.)

A measure $\nu$ on $(E,\salg E)$ is \emph{finite} if $\nu(E)<\infty$ and  \emph{$\sigma$-finite} if there exist a countable collection of pairwise disjoint sets $(E_i)_{i=0}^\infty\subset \salg B $ such that $E= \bigcup_{i=0}^\infty E_i$ and $\nu(E_i)<\infty$ for all $i\in \Nn$; clearly finite implies $\sigma$-finite.  The measure $\nu$ is a \emph{probability} measure if $\nu(E)=1$; this entails that $\nu(A)\leq 1$ for any $A\in \salg E$. When $\nu$ is a probability, the measure space $(E,\salg E,\nu)$ is also called a \emph{probability triple}.

Given a measure $\nu$ on $(E,\salg E)$, an event $A$ is said to occur \emph{$\nu$-almost everywhere (a.e.)} if $\nu(A)=1$; hen $\nu$ is a probability measure, the terminology \emph{$\nu$-almost surely (a.s.)} is used instead. If $\nu(A)=0$ the event $A$ is \emph{$\nu$-null}.

Given measurable spaces $(E,\salg E)$ and $(F,\salg F)$, a function $f:E\to F$ is called \emph{measurable} if $f^{-1}(F)\in \salg E$ for all $F\in \salg F$. If $(E,\salg E)$ is equipped with a measure $\mu$, $f$ induces a measure $f_*\mu$ on $(F,\salg F)$ called \emph{image measure} or \emph{push-forward measure}, defined by
\begin{equation}
\forall F\in \salg F, \quad (f_*\mu)(F)= \mu(f^{-1}(F)).
\end{equation}

Given any collection $\salg C$ of subset of $F$, $\sigma(\salg C)$ denotes the smallest $\sigma$-algebra that contains $\salg C$. To verify that a function $f:E\to F$ from $(E,\salg E)$ to $(F,\sigma(\salg C))$ is measurable, it is enough to verify that $f^{-1}(C)\in \salg E$ for all $C\in \salg C$.

If $(X,\tau)$ is a topological space,  $\sigma(\tau)$ is called the \emph{Borel $\sigma$-algebra} and its elements are called \emph{Borel sets}. In all the examples we have in mind the $\sigma$-algebras arise in this way. In many situations the Euclidean space is treated as a measurable space; it is implicit that it is equipped with the Borel $\sigma$-algebra. In virtue of the previous paragraph, to verify that a real-valued function $f:E\to \Rr^d$ defined on some measure space $(E,\salg E)$ is measurable, it is enough to show that $f^{-1}(A)\in \salg E$ for any open set $A$ or even for any $A$ in a sub-basis of the topology (e.g. the intervals of the form $(-\infty, a)$ or $(b,\infty)$ when $d=1$). 

Given measure spaces $(F,\salg F)$ and $(G,\salg G)$, the product $\sigma$-algebra is 
\begin{equation}
\salg F\otimes \salg G =\sigma(\set{A\times B}{A\in \salg F\,B\in \salg G}).
\end{equation}
We abbreviate $\salg F\otimes \salg F=\salg F^{\otimes 2}$. The $n$-fold product $\salg F^{\otimes n}$ is defined by iteration of this construction. Given $\sigma$-finite measures $\nu$ and $\xi$ respectively on $(F,\salg F)$ and $(G,\salg G)$, $\nu \otimes \xi$ is the only measure on $\salg F\otimes \salg G$  that satisfies $(\nu \otimes \xi)(A\times B)=\nu(A)\xi(B)$ for any $A\in \salg F$ and $B\in \salg G$ \cite[Thm. 5.1.4]{Cohn2013}.

For us, a \emph{random variable} is any measurable function $X:\Omega\to E_X$, from some probability triple $(\Omega,\salg F, \nu)$ to a measurable set $(E_X,\salg B)$. In this case the \emph{law} of $X$ is the push-forward measure $X_*\nu$. (Usual textbooks in probability theory, such as \cite{Williams1991}, deal mainly with real-valued random variables, but here we accept more general codomains.)

A function $s:E\to \Rr$ on a measure space $(E,\salg E,\mu)$ is called \emph{simple} is it is of the form $s=\sum_{i=1}^N a_i \cf_{A_i}$ for some $N\in \Nn$, $(a_1,...,a_n)\in \Rr^N$ and $\{A_1,..,A_N\}\subset \salg E$. Its integral is, by definition, $ \int_E s\d\mu = \sum_{i=1}^N a_i \mu(A_i)$. The Lebesgue integral of a positive measurable function $p:E\to [0,\infty)$  is defined as
\begin{equation}
\int_E p\d\mu = \sup\set{\int_E s\d\mu}{s \text{ simple and }s\leq p}.
\end{equation}
Such a $p$ is called \emph{integrable} if this supremum is finite. A signed measurable function $f$ is first decomposed into its positive and negative parts 
\begin{equation}
f = f^+ - f^-, \quad f^+=\max(f,0),\quad f^{-} = \max(-f,0),
\end{equation}
and called \emph{integrable} if $\int_Ef^{+}\d\mu <\infty$ and $ \int_E f^{-1}\d\mu <\infty$; in such a case, $\int_E f\d\mu := \int_Ef^{+}\d\mu - \int_E f^{-1}\d\mu$. Sometimes we  write instead $\int_E f(x) \d\mu(x)$; the meaning is the same. If $A\in \salg E$, then $\int_A f\d\mu := \int_E f\cf_A \d\mu$; the map $f\mapsto \int_A f\d\mu$ is linear.

Given two measures $\nu$ and $\xi$ on $(E,\salg E)$, we say that $\nu$ is absolutely continuous with respect to $\xi$ (denoted $\nu\ll\xi$) if for all $A\in \salg E$, $\xi(A)=0$ implies $\nu(A)=0$.  This is the case if $\nu(A) = \int_A f\d\xi$ for all $A\in \salg E$.  Conversely, the Radon-Nikodym theorem \cite[Thm. 4.2.2]{Cohn2013} says that, given $\sigma$-finite measures $\nu$ and $\xi$ on $(E,\salg E)$, if $\nu \ll \xi$ then there exists an integrable function $f:E\to \Rr$ such that 
\begin{equation}\label{eq:condition_radon_nikodym}
\forall A\in \salg E, \quad \nu(A) = \int_A f\d\xi,
\end{equation}
that is unique in the following sense: if $f'$ also satisfies \eqref{eq:condition_radon_nikodym}, then $\xi(\{f\neq f'\})=0$ (in that case, one says that $f$ and $f'$ define the same class in $L^1(E,\salg E,\xi)$).

\subsection{Outer measures and Hausdorff measure}

To construct a measure, one can follow a procedure attributed to Carath\'eodory \cite[p. 54]{Federer1969}. For this, one introduces first an \emph{outer measure} $\nu^*$ on a set $E$, that is, a $\nu^*:2^{E}\to [0,\infty]$ that satisfies
\begin{enumerate}
\item $\nu^*(\emptyset)=0$;
\item if $A\subset B \subset E$, then $\nu^*(A)\leq \nu^*(B)$, and
\item for any countable collection $(A_i)_{i\in \Nn}$ of subsets of $E$,
$$ \nu^*\left(\bigcup_{i\in \Nn} A_i\right) \leq \sum_{i\in \Nn} \nu^*(A_i).$$
\end{enumerate}
Then a set $A\subset E$ is called \emph{$\nu^*$-measurable} if 
\begin{equation}
\forall T\subset E, \quad \nu^*(T) = \nu^*(T\cap A) + \nu^*(T\setminus A).
\end{equation}
One can prove that the collection $\salg M_{\nu^*}$ of all $\nu^{*}$-measurable subsets is a $\sigma$-algebra and that the restriction of $\nu^*$ to $\salg M_{\nu^*}$ is a measure \cite[Thm. 1.3.6]{Cohn2013}, \cite[Thm. 2.1.3]{Federer1969}. 

\begin{example}[Counting measure]
The function $\nu^\#:2^{E}\to [0,\infty]$ that maps a set $A\subset E$ to the number of elements it contains (possibly $\infty$) is an outer measure. All subsets of $E$ are $\nu^{\#}$-measurable \cite[p. 54]{Federer1969}. 
\end{example}

The following example, arguably the most important for our purposes, is one of the central constructions of Geometric Measure Theory.

\begin{example}[Hausdorff measure]
The diameter of a subset $S$ of the Euclidean space $\Rr^d$ is  
 \begin{equation}
 \diam(S) = \sup\set{\norm{x-y}_2}{x,y\in S}.
 \end{equation}
 For any $m\geq 0$ and any  $A\subset \Rr^n$, define 
\begin{equation}\label{eq:Hausdorff_meas}
\mathcal H^m(A) = \lim_{\delta\to 0} \inf_{\{S_i\}_{i\in I}} \sum_{i\in I} a_m \left(\frac {\diam(S_i)}{2}\right)^m,
\end{equation}
where $a_m$ is a constant and the infimum taken over all countable coverings $\{S_i\}_{i\in I}$ of $A$ such that each set $S_i$ has diameter at most $\delta$.  This is an outer measure\footnote{In \cite[Def. 1.48]{Ambrosio2000}, outer measures are only defined for metric spaces and are subject to the additional condition
$$\forall E,F\subset S, \quad \operatorname{dist}(E,F) \Rightarrow \nu(E\cup F) = \nu(E) + \nu(F).$$
Remark that the Hausdorff measure is also an outer measure in this more restrictive sense.}
 on $\Rr^d$ for any $a_m>0$; its restriction to the $\mathcal H^m$-measurable sets is also denoted $\mathcal H^m$ and called \emph{$m$-Hausdorff measure.} All Borel sets are $\mathcal H^m$-measurable \cite[Thm. 1.49]{Ambrosio2000}, so the restricted set-function $\mathcal H^m:\salg B(\Rr^d) \to [0,\infty]$ is a measure. We further suppose that  $a_m=w_m:= \pi^{m/2}/\Gamma(m/2+1)$, which coincides with the volume of the ball $B(0,1)\subset \Rr^m$ when $m$ is an integer, so that the measure $\mathcal H^d$ equals the standard Lebesgue measure $\lambda^d$ \cite[Thm. 2.53]{Ambrosio2000}. For simplicity, the reader might take this as the definition of the standard Lebesgue measure, although it  is usually defined using coverings by ``rectangles'' $[a_1,b_1]\times \cdots \times [a_n,b_n]$ and not general sets. In turn, $\mathcal H^0$ equals the counting measure $\nu^{\#}$.
 
 For a gentle introduction to the Hausdorff measure and related topics, see \cite{Morgan2008}. 
\end{example}

We shall use the following lemma repeatedly.

\begin{lemma} \label{lem:inequality_hausdorff_Lipschitz_maps}
If $f:\Rr^d\to \Rr^{d'}$ is a Lipschitz function,\footnote{A function $f:X\to Y$ between metric spaces $(X, d_X)$ and $(Y,d_Y)$ is called Lipschitz if there exists an $L>0$ such that
$$\forall x,x'\in X,\quad d_Y(f(x),f(x')) \leq L d_X(x,x').$$
The Lipschitz constant  of $f$, denoted sometimes $\operatorname{Lip}(f)$, is the smallest $L$ that satisfies this condition.
} 
with Lipschitz constant $L$, then 
\begin{equation}\label{eq:inequality_hausdorff_Lipschitz_maps}
\forall m\geq 0 ,\,\forall E\subset \Rr^d,\quad \mathcal H^m(f(E)) \leq L^k \mathcal H^m(E).
\end{equation}
\end{lemma}
\begin{IEEEproof}
This follows easily from the inequality $\diam(f(E))^k \leq L^k \diam(E)^k$ \cite[Prop.~2.49]{Ambrosio2000}.
\end{IEEEproof}

 It is also worth mentioning here that in most references on Geometric Measure Theory e.g. Federer's book \cite{Federer1969} and Mattila's book \cite{Mattila1995}, the word \emph{measure} refers to an outer measure. An outer measure on a topological space is called Borel if all Borel sets are measurable. An outer measure $\nu$ on a locally compact Hausdorff space $X$ is a Radon outer measure if it is a Borel outer measure and it   satisfies some regularity conditions: 
\begin{enumerate}
\item $\nu(K)<\infty$ for all compact sets $K$,
\item $\nu(V) = \sup \set{\nu(K)}{K\subset V, \, K \text{ compact}}$ for every open $V\subset X$,
\item $\nu(A) = \inf\set{\nu(V)}{A\subset V, \, V \text{ open}}$ for any $A\subset X.$
\end{enumerate}
This terminology will be relevant in the definition of rectifiable measures (Section \ref{sec:rectifiable_measures}).

\subsection{Asymptotic equipartition property}

Now that we have the language of measure theory at our disposal, we can state the asymptotic equipartition property in the greatest possible generality. The only assumptions are those required to invoke the Radon-Nikodym theorem and the  law of large numbers.

Let  $(E_X, \salg B)$ be a measure space, that we conceive as the codomain of some random variable $X$, and let $\mu$ be a $\sigma$-finite measure on  $(E_X, \salg B)$ that encodes a relevant notion of ``size'' or ``volume.'' Finally, let  $\rho$ be a probability measure on  $(E_X, \salg B)$ absolutely continuous with respect to $\mu$, and  $f$ be a version of the Radon-Nikodym derivative $\d{\rho}/\mathrm{d}{\mu}\in L^1(E_X,\mu_X)$, which should be interpreted as a ``probability density function''. The \emph{generalized  entropy} of $\rho$ w.r.t. $\mu$ is defined as
\begin{equation}\label{eq:def_gen_entropy}
 \ent\mu(\rho):= -\int_{E_X} f(x) \ln f(x) \d\mu(x).
\end{equation}
This was introduced by Csisz\'ar in \cite{Csiszar1973}, see also Eq. (8) in \cite{Koliander2016}. Remark that the set where $f=0$, hence $\ln(f) = -\infty$, is $\rho$-negligible. In this article we refer to $H_{\mu}(\rho)$ simply as the entropy of $\rho$ w.r.t. $\mu$.

Let $\{X_i:(\Omega, \salg F, \mathbb P) \to (E_X, \salg B, \mu) \}_{i\in \Nn}$ be a collection of i.i.d random variables with law $\rho$. The density of the joint variable $(X_1,...,X_n) $ w.r.t. $\mu^{\otimes n}$ is given by  $f_{X_1,...,X_n} (x_1,...,x_n) = \prod_{i=1}^n f(x_i)$; by the remarks in Section \ref{sec:measure_theory}, it is enough to verify this on sets of the form $A_1\times \cdots \times A_n$. If $f\log f$ is integrable, that is, if the Lebesgue integral  in \eqref{eq:def_gen_entropy}
is finite, then 
\begin{equation}
-\frac{1}{n} \ln f_{X_1,...,X_n}(X_1,...,X_n) \to \ent\mu(\rho)
\end{equation}
$\mathbb P$-almost surely (resp. in probability) as a consequence of the strong (resp. weak) law of large numbers, which is applied to the $[-\infty,\infty]$-valued random variables $\{\log f(X_i)\}_{i\in \Nn}$. The convergence in probability is enough to establish the following result.

\begin{proposition}[Asymptotic Equipartition Property]\label{prop:AEP}
Let $(E_X,\salg B,\mu)$ be a $\sigma$-finite measure space, and $\rho$ a probability measure on $(E_X,\salg B)$ such that  $\rho \ll \mu$ and $\ent\mu(\rho)$ is finite. For every $\delta >0$, define $W_\delta^{(n)}(\rho;\mu) $ as the set of points $(x_1,...,x_n)\in E_X^n $ that satisfy 
\begin{equation*}
\left|-\frac 1n  \ln f_{X_1,...,X_n}(X_1,...,X_n) - \ent\mu(\rho)\right| \leq \delta.
\end{equation*}
Then,
\begin{enumerate}
\item\label{AEP1} for every $\epsilon>0$, there exists $n_0\in \Nn$ such that, for all $n\geq n_0$, $$\Pr{W_\delta^{(n)}(\rho;\mu)}>1-\epsilon;$$
\item\label{AEP2} for every $n\in \Nn$, $$\mu^{\otimes n}(W_\delta^{(n)}(\rho;\mu)) \leq \exp\{n(\ent\mu(\rho)+\delta)\};$$
\item\label{AEP3} for every $\epsilon>0$, there exists $n_0\in \Nn$ such that, for all $n\geq n_0$, $$\mu^{\otimes n}(W_\delta^{(n)}(\rho;\mu)) \geq (1-\epsilon)  \exp\{n(\ent\mu(\rho)-\delta)\}.$$
\end{enumerate}
\end{proposition}

We proved these claims in \cite[Ch.~12]{Vigneaux2019-thesis}; our proofs are very similar to the standard ones for (Euclidean) differential entropy, see \cite[Ch.~8]{Cover2006}.

The elements of $W_\delta^{(n)}(\rho;\mu)$ are called weakly typical sequences. Below, we write $W_\delta^{(n)}(\rho)$ or even $W_\delta^{(n)}$ if $\rho$ and $\mu$ are clear from context. The choice of a version of $\d \rho/\d \mu$ is always assumed. 

When $E_X$ is a countable set and $\mu$ the counting measure, every probability law $\rho$ on $E_X$ is  absolutely continuous with respect to $\mu$; if $p:E_X\to \Rr$ is its density, $\ent{\mu}(\rho)$ corresponds to the familiar expression $-\sum_{x\in E_X} p(x)\ln p(x)$. 

If $E_X= \Rr^n$, $\mu$ is the corresponding Lebesgue measure, and $\rho$ a probability law such that $\rho \ll \mu$, then  the derivative $\d \rho/\d\mu \in L^1(\Rr^n)$ corresponds to the elementary notion of density, and the quantity   $\ent\mu(\rho)$ is the \emph{differential entropy} that was also introduced by Shannon in \cite{Shannon1948}. He remarked that the covariance of the differential entropy under diffeomorphisms is consistent with the measurement of randomness ``\emph{relative to an assumed standard}.'' For example, consider a linear automorphism of $\Rr^n$,  $\varphi(x_1,...,x_n)= (y_1,...,y_n)$,  represented by a matrix $A$. Set $\mu = \d x_1 \cdots \d x_n$ and $\nu = \d y_1 \cdots \d y_n$. It can be easily deduced from the change-of-variables formula that $\nu(\varphi(V)) = |\det A| \mu(V)$. Similarly,  $\varphi_*\rho$ has density $f(\varphi^{-1}(y))|\det A|^{-1}$ w.r.t. $\nu$, and this implies that $\ent{\nu}(\varphi_*\rho) = \ent{\mu}(\rho) + \ln |\det A|$, cf. \cite[Eq.~8.71]{Cover2006}. Hence
\begin{multline}
\left| -\frac 1n \ln \prod_{i=1}^n \diff{\varphi_*\rho}{\nu}(y_i) - \ent{\nu}(\varphi_*\rho)\right| = \\ \left| -\frac 1n \ln \prod_{i=1}^n \diff{\rho}{\mu}(\varphi^{-1}(y_i)) - \ent{\mu}(\rho)\right|,
\end{multline}
from which we deduce that $W_{\delta}^{(n)}(\varphi_* \rho, \nu) = \varphi^{\times n}(W_\delta^{(n)}(\rho;\mu))$ and consequently 
\begin{equation}
\nu^{\otimes n} (W_{\delta}^{(n)} (\varphi_*\rho; \nu)) = |\det A|^n \mu^{\otimes n}(W_{\delta}^{(n)}(\rho; \mu)),
\end{equation}
which is  consistent with the corresponding estimates given by Proposition \ref{prop:AEP}.

In the discrete case one could also work with any multiple of the counting measure, $\nu = \alpha \mu$, for $\alpha>0$. In this case, the chain rule for Radon-Nikodym derivatives (see  \cite[Sec.~19.40]{Hewitt1965}) gives
\begin{equation}
\diff{\rho}{\mu}= \diff{\rho}{\nu}\diff{\nu}{\mu} = \alpha \diff{\rho}{\nu},
\end{equation}
and therefore $
\ent{\mu}(\rho) =\ent{\nu}(\rho) - \ln\alpha.$
Hence the discrete entropy depends on the choice of reference measure, contrary to what is usually stated. This function is invariant under a bijection of finite sets, but taking on both sides the counting measure as reference measure. The proper analogue of this in the Euclidean case is a measure-preserving transformation (e.g. $|\det A|=1$ above), under which the differential entropy \emph{is} invariant.

For any $E_X$, if $\mu$ is  a probability law, the expression $\ent\mu(\rho)$ is the opposite of the \emph{Kullback-Leibler divergence} $D_{KL}(\rho||\mu) := -\ent{\mu}(\rho).$ The positivity of the divergence follows from a customary application of Jensen's inequality.

The asymptotic relationship between volume and entropy given by the AEP can be summarized as follows:

\begin{corollary}\label{cor:entropy_and_volume}
$$\lim_{\delta \to 0} \lim_{n\to \infty} \frac{1}{n}\ln \mu^{\otimes n}(W_\delta^{(n)}(\rho;\mu))=\ent{\mu}(\rho).$$
\end{corollary}

\subsection{Rectifiable sets }\label{sec:rectifiable_sets}

In \cite{Koliander2016}, Riegler, Hlawatsch, Koliander, and Pichler introduced another family of examples for the AEP as stated in the previous section. In their work, the space $E_X$ is an $(\mathcal H^m,m)$-rectifiable subset of $\Rr^d$, and the reference measure $\mu$ equals $\mathcal H^m\mres E_X$,  the restriction of the $m$-dimensional Hausdorff measure $\mathcal H^m$  to $E_X$, which turns out to be $\sigma$-finite.\footnote{Given a measure $\mu$ and a $\mu$-measureble set $A$, the restriction of $\mu$ to $A$ is the measure $\mu\mres A$ given by $(\mu\mres A)(B):=\mu(A\cap B)$.}

Let $S$ be a Borel subset of $\Rr^d$, and let $m$ be its Hausdorff dimension \cite[Def. 2.51]{Ambrosio2000}
\begin{align}
m&= \inf\set{k\in[0,\infty)}{\mathcal H^k (S)=0} \\
&= \sup\set{k\in[0,\infty)}{\mathcal H^k (S) = \infty}.
\end{align}
 The set $S$ can be seen as a measurable space with the induced $\sigma$-algebra $\salg B(S) := \set{S\cap B}{B\in \mathcal B(\Rr^d)}$, which is also the Borel $\sigma$-algebra for the induced topology on $S$. The restricted measure $\mathcal H^m \mres S$ can be seen as a measure on $\salg B(\Rr^d)$ or $\salg B(S)$ depending on the context. We introduce now a particular family of subsets $S$ such that $\mathcal H^m\mres S$ behaves well; they also play a fundamental role in Geometric Measure Theory.

\begin{definition}[Rectifiable sets]
A subset $S$ of $\mathcal \Rr^d$ is called \emph{$m$-rectifiable} (for $m\leq d$) if it is the image of a bounded subset of $\Rr^m$ under a Lipschitz map and \emph{countably $m$-rectifiable set} if it is a countable union of $m$-rectifiable sets.  The subset $S$ is  
 \keyt{countably $(\mathcal H^m,m)$-rectifiable}  if there exist countable $m$-rectifiable set containing $\mathcal H^m$-almost all of $S$, this is, if there are bounded sets $A_k \subset \Rr^m$ and Lipschitz functions $f_k : A_k \to \Rr^d$, enumerated by $k\in \Nn$, such that $\mathcal H^m(S\setminus \bigcup_k f_k(A_k))=0$. Finally, the subset $S$ is called $(\mathcal H^m,m)$-rectifiable if it is countably $(\mathcal H^m,m)$-rectifiable and $\mathcal H^m(S) < \infty$. 
\end{definition}

By convention, $\Rr^0$ is a point, so  that a countably $0$-rectifiable set is simply a countable set. 

\begin{example}\label{ex:manifolds_rectifiable}
A subset $E$ of $\Rr^d$ is an $m$-dimensional $C^1$-submanifold of $\Rr^d$ \cite[Def. 5.3.1]{Krantz2008} if every point $x\in E$ has an open neighborhood $V$ in $\Rr^d$ such that there exists a bijective $C^1$ map $\varphi: U\to \Rr^d$, for some open $U\subset \Rr^m$, satisfying 
\begin{enumerate}
\item $D\varphi$ has rank $m$ at all points of $U$, and
\item $\varphi(U) = V\cap E$.
\end{enumerate}
The function $\varphi$ is called a coordinate map. The definition implies that $E$ is $m$-rectifiable: for any $x\in E$, there is coordinate map $\varphi:U\to \Rr^m$ with image $V\cap E$ containing $x$ and we can introduce  a compact ball $K_{\varphi^{-1}(x)}$ centered at $\varphi^{-1}(x)$; as a consequence of the Mean Value Theorem, $\varphi$ is Lipschitz on $K_{\varphi^{-1}(x)}$ with constant $\sup_{y\in K_{\varphi^{-1}(x)}} \norm{\nabla \varphi(y)}$. Usual topological arguments, relying on the fact that every cover of an open subset of $\Rr^d$ has a countable subcover (Lindel\"of property), show that $E$ is a countable union of sets $\{\varphi^{-1}(K_{\varphi(x_i)})\}_i$, see Appendix A in \cite{Alberti2019}
\end{example}

\begin{remark}\label{rmk:properties_H_measure}
It follows from Lemma \ref{lem:inequality_hausdorff_Lipschitz_maps} and the boundedness of the sets $A_k$ that
every $(\mathcal H^m,m)$-rectifiable set has $\sigma$-finite $\mathcal H^m$-measure. Because of monotonicity of $\mathcal H^m$,  any subset of an $(\mathcal H^m,m)$-rectifiable set is $(\mathcal H^m,m)$-rectifiable. Finally, remark that the countable union of $(\mathcal H^m,m)$-rectifiable subsets is $(\mathcal H^m,m)$-rectifiable \cite[Lem.~15.5]{Mattila1995}. 
\end{remark}

\begin{remark}
Given that we use many results from Federer's book \cite{Federer1969}, we prefer to stick to his terminology. In Mattila's book \cite{Mattila1995}, another standard references about rectifiable sets in $\Rr^d$, countably $(\mathcal H^m,m)$-rectifiable sets are called $m$-rectifiable. Something akin to a countably $(\mathcal H^m,m)$-rectifiable set is called  $m$-rectifiable in \cite{Koliander2016}, although they also assume that each $A_k$ is Lebesgue measurable (which makes their use of Federer's results less evident).
\end{remark}

It turns out that a subset $S$ of $\Rr^d$ is  countably $(\mathcal H^m,m)$-rectifiable if and only if it is contained in the union of an $\mathcal H^m$-null set and countably many $m$-dimensional $C^1$-submanifolds of $\Rr^d$ \cite[Lem. 5.4.2]{Krantz2008}. For our purposes,  the following representation of $\mathcal H^m$-measurable  $(\mathcal H^m, m)$-rectifiable sets \cite[Lem.~3.2.18]{Federer1969} will be preferable.

\begin{lemma}\label{lem:compact_decomposition} If $W$ is $(\mathcal H^m, m)$-rectifiable and $\mathcal H^m$-measurable subset of $\Rr^d$, and if $1 < \ell < \infty$, there exist compact subsets $K_1$, $K_2$, $K_3, \cdots$ of $\Rr^m$ and Lipschitz functions $f_1$, $f_2$, $f_3,\cdots$ of $\Rr^m$ into $\Rr^d$ such that $f_1(K_1)$, $f_2(K_2)$, $f_3(K_3),\cdots$ are pairwise disjoint subsets of $W$ with 
$$\mathcal H^m\left( W \setminus \bigcup_{i=1}^\infty f_i(K_i)\right) = 0$$
and, for each $i\geq 1$, $\Lip(f_i) \leq \ell$, $ f_i|_{K_i}$ is injective,\footnote{Federer uses the term \emph{univalent}.} and $ \Lip((f_i|_{K_i})^{-1})\leq \ell.$
\end{lemma}

In other words, the lemma says that says that
\begin{equation}\label{eq:rep_rectifiable}
W\subset W_0 \cup \bigcup_{i\in \Nn} f_i(K_i),
\end{equation}
where $W_0$ is an $\mathcal H^m$-measurable, $\mathcal H^m$-null set and each $f_i(K_i)$ is a Borel set, being the image of a Borel set under an injective map.\footnote{\textbf{Theorem (Lusin-Souslin):} Let $X,Y$ be Polish spaces and $f:X \to Y$ be continuous. If $A \subset X$ is Borel and $f|_A$ is injective, then $f(A)$ is Borel \cite[Thm.~15.1]{Kechris2012}. \label{footnote_Lusin_souslin}}  This formula also holds for countably $(\mathcal H^m,m)$-rectifiable sets. Compare with  Remark 5 in \cite{Koliander2016}. 

In virtue of Lemma \ref{lem:inequality_hausdorff_Lipschitz_maps},
\begin{equation}
\mathcal H^m(f_i(K_i)) \leq (\Lip(f))^m\mathcal H^m(K_i),
\end{equation}
and evidently  $\mathcal H^m(K_i)= \lambda^m(K_i)< \infty$. Hence  each set $f_i(K_i)$ has finite $m$-Hausdorff; we can suppose this measure to be strictly positive without loss of generality. We conclude that the Hausdorff dimension of each set $K_i$ and $f_i(K_i)$ is then $m$.

\section{Piecewise continuous measures}\label{sec:piecewise_continuous}

\subsection{Definition}\label{sec:piecewise_cont_defi}

Although in this article we are mainly interested in stratified measures, many results hold in the more general framework of piecewise continuous measures. 

\begin{definition}[Piecewise continuous measure] \label{defi:piecewise_continuous}
A measure $\rho$ on $(E_X, \salg B)$ is \emph{piecewise (absolutely) continuous} if, for some $k\in \Nn^*$,
there exists a finite collection of pair-wise disjoint sets $E_1,...,E_k \in \salg B$ and  $\sigma$-finite measures $(\mu_i)_{i=1}^k$ on $(E_X,  \salg B)$ such that 
\begin{enumerate}
\item $\mu_i = \mu_i\mres E_i$,
\item $\rho(E_i)>0$ for all $i=1,..,n$,
\item  $\rho(E_X\setminus \bigcup_{i=1}^k E_i)=0$, and 
\item $\tilde \rho_i := \rho\mres E_i \ll \mu_i$.
\end{enumerate}
\end{definition}

Henceforth we suppose, without real loss of generality, that $E_X = \bigcup_{i=1}^kE_i$ for each $i=1,...,k$. The measure $\mu:=  \sum_{i=1}^k \mu_i$ on $E_X$ is $\sigma$-finite, and the collection of measures $\{\mu_i\}_{i=1}^k$ give a disintegration of the measure $\mu$ under the surjection $\pi:E_X\to E_Y:= \{1,...,k\}$ that maps $E_i$ to $i$.  (We could also allow $E_0:=E_X\setminus \bigcup_{i=1}^kE_i$ to be nonempty, but then we would need to introduce an additional symbol in $E_Y$. The notation would be more complicated but the results would not be change, as the reader will see, because $E_0$ is $\rho$-null.)
 
 \begin{remark}[Disintegrations]
Let $(Z,\salg B_Z)$ and $(W, \salg B_W)$ be measure spaces, $\nu$ a measure on $(W, \salg B_W)$, and  $T:Z\to W$ a measurable map. A measure $\alpha$ on $Z$ has a $(T, \nu)$-disintegration if there exists a finite collection of measures $\{\alpha_{w}\}_{w\in W}$ such that:
\begin{enumerate}
\item for each $w\in W$,  $\alpha_w$ is a $\sigma$-finite measure on $\salg B_W$ concentrated on $Z_w := T^{-1}(w)$ i.e. $\alpha_w(Z\setminus Z_w)=0$;
\item for each measurable nonnegative function $f:Z\to \Rr$, the map $w\mapsto \int_E f(z) \d \alpha_w(z)$ is $\salg B_W$ measurable and
\begin{equation}
\int_E f(z) \d \alpha(z) = \int_W \int_Z f(z) \d \alpha_w(z) \d\nu(w).
\end{equation}
\end{enumerate}

Hence the measures $\{\mu_i\}_{i\in E_Y}$ give a $(\pi, \cmeas)$-disintegration of the measure the measure $\mu$ on $E_X$, where $\cmeas$ denotes the counting measure. Moreover, any probability measure absolutely continuous with respect to $\mu$ also has a disintegration \cite[Thm. 3]{Chang1997}. The piecewise continuous measure $\rho= \sum_{i=1}^k \tilde \rho_i$ is absolutely continuous with respect to $\mu$ and its $(\pi, \cmeas)$-disintegration is the family $\{\tilde \rho_i\}_{i\in E_Y}$. 
\end{remark}

A piecewise continuous \emph{finite} measure $\rho$ equals $\sum_{i=1}^k \tilde \rho_i =  \sum_{i=1}^k q_i  \rho_i$, where $q_i := \tilde \rho_i(E_i)=\rho(E_i)$ and $\rho_i = \tilde \rho_i/q_i$ is a probability measure for each $i=1,...,k$.  When $\rho$ is a probability measure, the vector $(q_1,...,q_k)$ defines a probability on the set $E_Y$; in fact, it is the probability mass function (p.m.f) of the image measure (push-forward) $\pi_*\rho$.

We give now two examples of piecewise continuous measures.

\begin{example}
Every absolutely continuous measure $\rho$ with respect to an arbitrary measure $\mu$ is piecewise continuous with respect to any finite measurable partition $\mathcal E$ of the ambient space. The sets $E_1,...,E_k$ are those $E\in \mathcal E$ such that $\rho(E)>0$ and each $\mu_i$ is defined as the restriction of $\mu$ to $E_i$.
\end{example}

\begin{example}\label{item:Lebesgue_decomposition}
Let $E_X$ be a locally compact group $G$ and $\mu$ a Haar measure on it. The Banach space of locally compact measures on $G$ decomposes as a direct sum, in such a way that any signed measure $\xi$ can be written uniquely as a sum
\begin{equation}
\xi = \xi_{ac} + \xi_{sc} + \xi_{pd},
\end{equation}
where $ \xi_{ac} \ll \mu$, the measure $\xi_{pd}$ is purely discrete, and $\nu_{sc}$ is singular continuous (it is supported on a $\mu$-null-set and all the singletons $\{x\}\subset G$ have $\xi_{sc}$-measure zero.), see \cite[Thm. 19.20]{Hewitt1994}. %By ``purely singular continuous'' we mean that $ \nu_{a.c.} =  \nu_{d} = 0$. 
Whenever $G$ is uncountable and $\xi_{sc}$ vanishes, the measure $\xi$ is piecewise continuous; to conform to the definition, one has to restrict $\xi_{ac}$ to $G\setminus S$, where $S$ is the  discrete support of $\xi_{pd}$, which has $\mu$-measure $0$. 
\end{example}

The stratified measures introduced in Section \ref{sec:stratified_def} generalize the preceding example, allowing a nontrivial singular continuous part.

\subsection{Chain rule: alternative expression for the entropy}\label{sec:chain_rule}

In this section, we consider a piecewise continuous \emph{probability} measure $\rho$ (as in Definition \ref{defi:piecewise_continuous}) written in \emph{mixture form} $\rho= \sum_{i=0}^k q_i \rho_i$, where $q_i=\rho(E_i)$ and $\rho_i =  (\rho\mres E_i)/q_i$. 

\begin{lemma}\label{lemma:density_piecewise}
Let $\rho$ be a piecewise continuous probability measure with mixture form $\rho = \sum_{i=1}^k q_i \rho_i$. The measure $\rho$ is absolutely continuous with respect to $\mu=\sum_{i=1}^k \mu_i$, and 
\begin{equation}
\diff{\rho}{\mu} = \sum_{i=1}^m q_i \diff{\rho_i}{\mu_i} \cf_{E_i}
\end{equation}
almost surely.
\end{lemma}
\begin{IEEEproof}
Let us prove first that $\rho \ll \mu$. For any $A\in \salg B$, $\mu(A) = 0$ implies that, for every $i=1,..,k$, $\mu_i(A) = 0$, because each $\mu_i$ is nonnegative. We deduce that $ \rho_i(A) = 0$, for each $i$, because $\rho_i\ll \mu_i$. Since $\rho  = \sum_i q_i \rho_i$, we also have $\rho(A)=0$.

Since the sets $E_1,...,E_n$ are disjoint, we have
\begin{equation}
\mu_i(B) = \mu_i(B\cap E_i) = \mu(B\cap E_i) = \int_B \cf_{E_i} \d \mu,
\end{equation}
 hence $\cf_{E_i} = \d \mu_i/\d\mu$.

Using now that $\rho_i \ll \mu_i \ll \mu$,
\begin{align}
\rho(A) &= \sum_{k=1}^m q_k \rho_k(A) \\
&= \sum_{k=1}^m q_k  \int_{A}   \diff{\rho_k}{\mu_k} \diff{\mu_k}{\mu} \d \mu  \\
&= \int_A \left( \sum_{k=1}^m q_k \cf_{E_k} \diff{\rho_k}{\mu_k} \right) \d \mu. 
\end{align}
The second line uses the chain rule of the Radon-Nikodym derivative, and the third follows from linearity of the integral. The proposition follows from uniqueness of the derivative in $L^1$. 
\end{IEEEproof}

\begin{lemma}[Chain rule]\label{lem:chain_rule}
For a piecewise continuous measure $\rho$ with mixture form $\sum_{i=1}^k q_i\rho_i$ (see above) the equality
\begin{equation}
\ent{\mu}(\rho) = \ent{\#}(q_1,...,q_n) + \sum_{i=1}^k q_i \ent{\mu_i}(\rho_i)
\end{equation}
holds.
\end{lemma}
%The term $\ent{}(p_1,...,p_m) := \ent{\cmeas}(\pi_*\rho) = -\sum_{k=1}^m p_k \log p_k$ is the familiar Shannon entropy.
\begin{IEEEproof}
We follow the usual convention $0\ln 0=$. In virtue of the previous lemma, 
\begin{align*}
-\int_{E_X} \diff{\rho}{\mu}\ln \diff{\rho}{\mu} \d\mu &= - \sum_{i=1}^k q_i \int_{E_i} \diff{\rho_i}{\mu_i} \ln\left(q_i \diff{\rho_i}{\mu_i} \right) \d\mu \\
&= - \sum_{i=1}^k q_i \int_{E_i} \left(\ln q_i + \ln \diff{\rho_i}{\mu_i} \right) \diff{\rho_i}{\mu_i}  \d\mu  \\
&= - \sum_{i=1}^k q_i \ln q_i \rho_i(E_i)  +  \sum_{i=1}^k q_i  \ent{\mu_i}(\rho_i),
\end{align*}
and $\rho_i(E_i)=1$ for every $i$.
\end{IEEEproof}

\begin{remark}
Let $\rho$ have a decomposition $q\rho_{ac} + (1-q)\rho_{pd}$ as in Example \ref{item:Lebesgue_decomposition} above. We denote by $\tilde \rho_{ac}$ the restriction of $\rho$ to $G\setminus S$, and similarly $\tilde \mu$ the restriction of $\mu$ to $G\setminus S$. Since $S$ is $\mu$-null, we have $\rho= q \tilde \rho_{ac} + (1-q) \rho_{pd}$, hence $\rho$ is a piecewise continuous measure. The preceding result reads
\begin{equation}
\ent\mu(\rho) = \ent\#(q,1-q) + q \ent{\tilde\mu}(\tilde  \rho_{ac}) + (1-q) \ent{\#}({\rho_{pd}}).
\end{equation}
Since 
\begin{equation}
\diff{\tilde  \rho_{ac}}{\tilde \mu} = \diff{\rho_{ac}}{\mu}
\end{equation}
on $G\setminus S$ and $S$ is $\mu$-null, we have
$\ent{\tilde\mu}(\tilde  \rho_{ac}) = \ent{\mu}( \rho_{ac}).$
In particular,  we get the identity \eqref{eq:entropy_simple_misture}.
\end{remark}
 
 \begin{remark}[Chain rule for general disintegrations]
 Let $T:(E_X,\salg B_X) \to (E_Y,\salg B_Y)$ be a measurable map between arbitrary measurable spaces, $\mu$ (respectively  $\nu$) a $\sigma$-finite measure on $(E_X,\salg B_X)$ (resp. $(E_Y,\salg B_Y)$), and $\{\mu_y\}$ a $(T, \nu)$-disintegration of $\mu$. Then any probability measure $\rho$ absolutely continuous w.r.t. $\mu$, with density $r$, has a $(T, \nu)$-disintegration $\{\tilde \rho_y\}_{y\in Y}$ such that $\rho_y \ll \mu_y$ with density $r|_{Y=y}$ a.s., 
 as well as  a $(T,T_*\rho)$-disintegration $\{ \rho_y\}_{y\in Y}$ such that, $T_*\rho$-almost surely, $\rho_y$ is a probability measure, whose density is $(\int_E r \d\mu_y)^{-1}r|_{Y=y}$ with respect to $\mu_y$ \cite[Thm. 3]{Chang1997}. The function $y \mapsto m_y(r):=\int_X r \d\mu_y$ is the density of $T_*\rho$ with respect to $\nu$.  Then,
 \begin{align}
 \ent{\mu}(\rho) =    \ent{\nu}(T_*\rho)+  \int_{E_Y}\ent{\mu_y}(\rho_y)\d T_*\rho(y).\label{chain_rule_dis}
 \end{align}
A piecewise continuous measure provides a particular case of the situation above, such that $E_Y$ is finite and $\nu$ is the counting measure $\cmeas$ (all the notation is consistent). Hence  Lemma \ref{lem:chain_rule} is just a particular case of \eqref{chain_rule_dis}. For details, see \cite{Vigneaux-GSI21-disintegrations}.
 \end{remark}

\subsection{Asymptotic interpretation of the conditional entropy}\label{sec:cond_entrop_asymp}

This section elucidates the  meaning of the conditional term $H(X|Y):= \sum_{i=1}^k q_i \ent{\mu_i}(\rho_i)$ in \eqref{lem:chain_rule} in terms of the AEP. From Corollary \ref{cor:entropy_and_volume}, we know that
\begin{equation}
H(X):=\ent{\mu}(\rho) = \lim_{\delta\to 0} \lim_{n\to \infty} \frac 1n \ln \mu^{\otimes n}(W_{\delta}^{(n)}(\rho))
\end{equation}
and
\begin{equation}
H(Y):=\ent{\#}(T_*\rho) = \lim_{\delta'\to 0} \lim_{n\to \infty} \frac 1n \ln |W_{\delta'}^{(n)}(T_*\rho)|.
\end{equation}
Combined with Lemma 2, the preceding equalities imply that
\begin{equation}\label{eq:limit_conditional_ent}
H(X|Y) = \lim_{\delta'\to 0} \lim_{n\to \infty} \ln\left( \frac{\mu^{\otimes n}(W_{\delta}^{(n)}(\rho))}{|W_{\delta'}^{(n)}(T_*\rho)|}\right),
\end{equation}
which constitutes a firs answer.

The projection $\pi:E_X\to E_Y$ induces a map $\pi^{\times n}: E_X^n \to E_Y^n$. The equation \eqref{eq:limit_conditional_ent}  suggests that $\exp(n H(X|Y))$ approximates the $\mu^{\otimes n}$-volume of $W_{\delta}^{(n)}(\rho) \cap (\pi^{\times n})^{-1}(\V y)$,  for $\V y \in W_{\delta}^{(n)}$, because asymptotically there are roughly $|W_{\delta'}^{(n)}(T_*\rho)|^{-1}$ of these typical $\V y$ and they are approximately equiprobable. The present section elaborates this idea. We state our results in slightly more general form, assuming that $\V y$ belongs to a set $A_{\delta'}^{(n)} \subset W_{\delta'}^{(n)}(\pi_*\rho)$ such that for any $\epsilon > 0$ one has $(\pi_*\rho)^{\otimes n}(A_{\delta'}^{(n)}) \geq 1-\epsilon$ if $n$ is greater than certain $n_0(\epsilon)\in \Nn$.  In other words, the $\V y$ are eventually subject to stronger conditions than $|-\frac 1n \ln \prod_{i=1}^n \diff{\pi_*\rho}{\cmeas}(y_i)-H(Y)|<\delta'$; they may be typical in a stronger sense. This will be convenient to get finer estimates for the dimension of the fibers $(\pi^{\times n})^{-1}(\V y)$ in the stratified case (Section \ref{sec:stratified}).

For every $\V y\in E_Y^n$, we introduce the \emph{ stratum} $\Sigma_{\V y}:=(\pi^{\times n})^{-1}(\V y)$. We  define the set of  \emph{(strongly) typical strata} as $\Sigma^{(n)}_{\delta'}:= (\pi^{\times n})^{-1}(A_{\delta'}^{(n)})$.

The elements of $T^{(n)}_{\delta,\delta'}:=\Sigma^{(n)}_{\delta'}\cap W_{\delta}^{(n)}(\rho)$ are called \emph{doubly typical} sequences (in $E_X^n$). The set $T^{(n)}_{\delta,\delta'}$ can be decomposed as the union of \emph{doubly typical strata} $T^{(n)}_{\delta,\delta'}(\V y) := (\pi^{\times n})^{-1}(\V y) \cap W_{\delta}^{(n)}(\rho)$, for $\V y \in A_{\delta'}^{(n)}$.

\begin{proposition}\label{prop:bounds_doubly_typical_realizations}
\begin{enumerate}
\item\label{doubly_typ_I}  $\mu^{\otimes n}(T^{(n)}_{\delta,\delta'}) \leq \e^{n(\ent{}(X)+\delta)}$.
\item\label{doubly_typ_II}  For every $\epsilon > 0$, there exists an $n_0\in \Nn$ such that for all $n\geq n_0$,
\begin{equation}\label{eq:proba_doubly_typical}
\rho^{\otimes n}(T^{(n)}_{\delta,\delta'})\geq 1-2\epsilon
\end{equation}
and
\begin{equation}\label{eq:lower-bound_volume_doubly_typ}
(1-2\epsilon) \e^{n(\ent{}(X)-\delta)} \leq \mu^{\otimes n}(T^{(n)}_{\delta,\delta'}) 
\end{equation}
\end{enumerate}
\end{proposition}
\begin{IEEEproof}
Part \eqref{doubly_typ_I} follows from the bounds $\rho^{\otimes n}(T^{(n)}_{\delta,\delta'})=\int_{T^{(n)}_{\delta,\delta'}} f \d\mu^{\otimes n} \leq 1$ and the fact that, on $W^{(n)}_\delta$ (hence also on $T^{(n)}_{\delta,\delta'}$), the density $f = \diff{\rho^{\otimes n}}{\mu^{\otimes n}}$ is lower-bounded by $\exp(-n(H_{\mu}(\rho)+\delta))$.

To prove \eqref{eq:proba_doubly_typical}, we apply the union bound:
\begin{align*}
\rho^{\otimes n}((\Sigma_{W}^{(n)})^c) &= \rho^{\otimes n}((\Sigma^{(n)})^c \cup (W_\delta^{(n)}(\rho))^c) \\
&\leq \rho^{\otimes n}((\Sigma^{(n)})^c) + \rho^{\otimes n}(W_\delta^{(n)}(\rho))^c)\\
&\leq 2\epsilon.
\end{align*}
Then the inequality \eqref{eq:lower-bound_volume_doubly_typ} follows from \eqref{eq:proba_doubly_typical} and the pointwise inequality $f\leq \exp(-n(H_{\mu}(\rho)-\delta))$ on $W^{(n)}_\delta$.
\end{IEEEproof}

The conditional entropy gives an upper-bound for the volume of each doubly typical stratum in the following sense.
\begin{proposition}\label{prop:upper_bound_vol_T} For any $\delta, \delta'>0$, $n\in \Nn$ and $\V y \in A_{\delta'}^{(n)}$,
\begin{equation}\label{eq:upper_bound_volume_fiber}
\mu^{\otimes n}( T^{(n)}_{\delta, \delta'} (\V y)) \leq \exp( n H(X|Y) + n (\delta + \delta ')) 
\end{equation}
\end{proposition}
\begin{IEEEproof}
Remark first that
\begin{equation}
\rho^{\otimes n}(T^{(n)}_{\delta,\delta'}(\V y)) \leq \rho^{\otimes n}(\Sigma_{\V y}) \leq \exp(-n ( \ent{\#}(\pi_*\rho)-\delta)).
\end{equation}
We can write
\begin{equation}
\rho^{\otimes n}(T^{(n)}_{\delta,\delta'}(\V y)) = \int_{T^{(n)}_{\delta,\delta'}(\V y)} f \d \mu^{\otimes n}
\end{equation}
where $f$ is the derivative of $\rho^{\otimes n}$ with respect to $\mu^{\otimes n}$. Finally observe that, on $W_{\delta}^{(n)}$ (and on $T^{(n)}_{\delta,\delta'}$), $f$ is lower bounded by $\exp(-n (\ent\mu(\rho) + \delta))$. Therefore 
\begin{equation*}
\int_{T^{(n)}_{\delta,\delta'}(\V y)} \e^{-n (\ent\mu(\rho) + \delta)}\d \mu^{\otimes n} \leq \exp(-n ( \ent{\#}(\pi_*\rho)-\delta)).
\end{equation*}
The result follows from a  rearrangement of the terms.
\end{IEEEproof}

Equivalently, \eqref{eq:upper_bound_volume_fiber} says that, for every $\V y\in A^{(n)}_{\delta'}$,
\begin{equation}
\frac 1n \ln \mu^{\otimes n}(T_{\delta,\delta'}(\V y)) \leq H(X|Y) + (\delta+\delta').
\end{equation}
This upper bound \eqref{eq:upper_bound_volume_fiber} is asymptotically tight in a sense made precise by the following proposition.

\begin{proposition}\label{prop:tightness_bound}
Set $\boldsymbol \delta =(\delta,\delta')$. For a fixed $\epsilon > 0$, let $B_{\epsilon}^{(n)}$ denote the subset of $A_{\delta'}^{(n)}$ made of sequences $\V y$ such that 
\begin{equation}\label{eq:def_Bn}
 \frac 1n \ln \mu^{\otimes n}(T_{\delta.\delta'}^{(n)}(\V y)) > H(X|Y)-\epsilon + \delta + \delta'.
\end{equation}
Then,
\begin{align}
\limsup_{\norm{\boldsymbol \delta} \to 0}\limsup_{n\to\infty} \frac 1n \ln |B_\epsilon^{(n)}| &= \lim_{\delta'\to 0}\lim_{n\to\infty} \frac 1n \ln |A_{\delta'}^{(n)}|\\
& = H(Y).
\end{align}
\end{proposition} 
\begin{IEEEproof}
Set $B^{(n)} := B^{(n)}_\epsilon$ and $C^{(n)}= A_{\delta'}^{(n)} \sm B^{(n)}$. By additivity, 
\begin{align}\label{eq:upper_bound_decomposed_T}
&\mu^{\otimes n}(T_{\delta,\delta'}^{(n)})= \sum_{\V y\in A_{\delta'}^{(n)}} \mu^{\otimes n}(T_{\delta,\delta'}^{(n)}(\V y)) \nonumber \\
&\leq 2\max\left\{\sum_{\V y\in B^{(n)}} \mu^{\otimes n}(T_{\delta,\delta'}^{(n)}(\V y)), \sum_{\V y\in C^{(n)}} \mu^{\otimes n}(T_{\delta,\delta'}^{(n)}(\V y))\right\}.
\end{align}
The cardinality of both $B^{(n)}$ and $C^{(n)}$ is upper-bounded by the cardinality of $A_{\delta'}^{(n)}$ which is in turn upper-bounded by $\exp\{n\left(H(Y) + \delta'\right)\}$. Combining this fact with the definition of $C^{(n)}$, we get
\begin{align}
\sum_{\V y\in C^{(n)}} & \mu^{\otimes n}(T_{\delta,\delta'}^{(n)}(\V y))\\ 
 &\leq \e^{n(H(Y) + \delta' ))}  \e^{n(H(X|Y) -\epsilon + (\delta + \delta'))}\\
&\leq \exp(n(H(X) -\epsilon + 2\delta'  + \delta )),
\end{align}
 We conclude from this that 
\begin{equation}\label{eq:growth_C}
\limsup_{\norm{\boldsymbol\delta}\to 0 }\limsup_{n\to \infty} \frac 1n \ln \left| \sum_{\V y\in C^{(n)}}  \mu^{\otimes n}(T_{\delta,\delta'}^{(n)}(\V y))\right| \leq H(X) -\epsilon.
\end{equation}

In turn, from \eqref{eq:upper_bound_volume_fiber},
\begin{equation}
\sum_{\V y\in B^{(n)}} \mu^{\otimes n}(T_{\delta,\delta'}^{(n)}(\V y)) \leq \e^{n(H(X|Y) +\delta+\delta')} |B^{(n)}|.
\end{equation}

Suppose now that $\limsup_{\norm{\boldsymbol \delta} \to 0} \limsup_{n\to \infty} \frac 1n \ln |B^{(n)}| := L < H(Y)$. We can deduce then, from \eqref{eq:upper_bound_decomposed_T} and \eqref{eq:growth_C}, that 
\begin{multline}
\limsup_{\norm{\boldsymbol \delta}\to 0}\limsup_{n\to\infty} \frac 1n \ln \mu^{\otimes n}(T_{\delta,\delta'}^{(n)})\\ \leq \max( H(X|Y)+L, H(X)-\epsilon)) < H(X).
\end{multline}
This contradicts \eqref{prop:bounds_doubly_typical_realizations}, which in particular implies
\begin{equation}
\ent{}(X) \leq \liminf_{\norm{\boldsymbol \delta}\to 0}\liminf_{n\to\infty} \frac 1n \ln \mu^{\otimes n}(T^{(n)}_{\delta,\delta'}).
\end{equation}

\end{IEEEproof}

\subsection{Strong typicality}\label{sec:strong_typicality}

In the previous subsection, the sets $A_{\delta}^{(n)}$ do not have an explicit description; we only require that $A_{\delta}^{(n)}\subset W_{\delta'}^{(n)}$ and that for any $\epsilon > 0$, there exists an $n_0\in \Nn$ such that $(\pi_*\rho)^{\otimes n}(A_{\delta'}^{(n)})\geq 1-\epsilon$ whenever $n\geq n_0$. Here we specialize our discussion to sets $A_{\delta'}^{(n)}$ given by \emph{strongly typical sequences} as defined by Csisz\'ar and K\"orner for the analysis of discrete memoryless sources \cite[Ch.~2]{Csiszar1981}. Their very explicit description will yield a finer result on the probability of doubly typical strata (Proposition \ref{prop:same_prob_for_given_type}) and strong dimensional estimates in Section \ref{sec:typical_dimension}.

Before introducing the definition, let us denote by $Q$ the p.m.f. of $\pi_*\rho$ and remark that a sequence $\V y\in E_Y^n$ induces a probability law $\tau_{\V y}$ on $E_Y$, known as \emph{empirical distribution} and  given by 
\begin{equation}
\tau_{\V y}(\{a\}) = N(a;\V y)/n.
\end{equation}
Remember that $N(a;\V y)$ is the number of appearances of $a$ in $\V y$.

\begin{definition}[Strong typicality]\label{defi:strong_typicality}
A sequence $\V y\in E_Y^n$ is strongly $(Q,\eta)$-typical if it induces an empirical distribution $\tau_{\V y}$ with p.m.f. $P$  such that  $\tau_{\V y} \ll \pi_*\rho$ (we also write $P\ll Q$) and, for all $a\in E_Y$, $|P(a)-Q(a)|<\eta$.
\end{definition}

 We say that $\V y$ is of \emph{type} $P$ or that $P$ is the type of $\V y$. The definition implies that, for a strongly $(Q,\eta)$-typical type $P$, the inequality $d_{TV}(P,Q) < |E_Y| \eta$ holds, where $d_{TV}$ denotes the total variation distance $d_{TV}(P,Q)= \sum_{a\in E_Y} |P(a)-Q(a)|$. 

It is known that the entropy is continuous with respect to the total variation distance. In fact, if $d_{TV}(P,Q)=\theta \leq \frac 12$ ,
\begin{equation}
|H_{\#}(P)-H_{\#}(Q)| \leq -\theta \ln \frac{\theta}{|E_Y|}.
\end{equation}
cf. \cite[Lem.~2.7]{Csiszar1981}. So in particular, when $P$ is $(Q,\eta)$-typical, we get the upper bound
\begin{equation}\label{eq:bound_ent_eta_typical}
|H_{\#}(P)-H_{\#}(Q)| \leq - |E_Y|\eta \ln \eta.
\end{equation}
Hence given any $\delta'>0$, one can choose an $\eta=\eta(\delta')$ such that $A_{\delta'}^{(n)}$ is indeed a subset of $W_{\delta'}^{(n)}$. 

As in \cite[p.~21]{Csiszar1981}, we adopt the following \emph{$\eta$-convention}:\footnote{In their book they use the letter $\delta$ where we use $\eta$.}  for each $n\in \Nn$, choose some $\eta_n$ in such a way that $\eta_n\to 0$ and $\sqrt{n}\eta_n \to \infty$. For concreteness, we shall suppose that $\eta_n = n^{-1/2 +\xi}$, for some $\xi > 0$. In this case, we will also suppose that $\delta'_n = -|E_Y|\eta_n\ln \eta_n$; remark that not only $\eta_n$ tends to zero as $n\to \infty$, but $\delta'_n$ too. 

We assume from this point that $A_{\delta'_n}^{(n)}$ (sometimes simply denoted $A_{\delta'}^{(n)}$) is the set of $(Q,\eta_n)$ typical sequences. It is important below that for all $n$,
\begin{equation}
(\pi_*\rho)^{\times n}(A_{\delta'_n}^{(n)})\geq 1-\epsilon_n,
\end{equation}
where $\epsilon_n = 2|E_Y|\e^{-2n\eta_n^2}$. This is an elementary  consequence of Hoeffding's inequality; see also \cite[Lem.~2.12]{Csiszar1981}.

Given a possible type $P$ of sequences in $E_Y^n$, the \emph{type class} $T^{(n)}_P$ is the set of all sequences in $E_Y$ that have such a type.

\subsection{Symmetry}\label{sec:symmetry}

We establish in this section that all the doubly typical strata associated with strongly typical realizations of $Y$ of a given type have the same probability. If the volume is finite, they also have the same volume. Then we sketch a plan to lower-bound the volume of a douby typical fiber in terms of $H(X|Y)$.

\begin{proposition}\label{prop:same_prob_for_given_type}
For every type $P$ and every $\V y, \V y' \in T_P^{(n)}$, 
\begin{equation}\label{eq:equality_measure_fibers_type}
\rho^{\otimes n} (T_{\delta,\delta'}^{(n)}(\V y)) = \rho^{\otimes n} (T_{\delta,\delta'}^{(n)}(\V y')).
\end{equation}
Moreover, if the law $\mu$ is finite,
\begin{equation}\label{eq:equality_measure_fibers_type}
\mu^{\otimes n} (T_{\delta,\delta'}^{(n)}(\V y)) = \mu^{\otimes n} (T_{\delta,\delta'}^{(n)}(\V y')).
\end{equation}
\end{proposition}

Before giving a proof, we study certain symmetries of the problem. The permutation group $\mathfrak S_n$ acts on $E_X^n$ and $E_Y^n$ by permuting the components, this is, $\sigma \cdot(x_1,...,x_n) = (x_{\sigma(1)},...,x_{\sigma(n)}).$ The projection $\pi^{\times n}$ is clearly $\mathfrak S_n$-equivariant:
\begin{equation}\label{eq:equivariance}
\forall x\in E_X^n, \,\forall \sigma\in \mathfrak S_n,\quad \pi^{\times n}(\sigma\cdot \V x) = \sigma \cdot \pi^{\times n}(\V x).
\end{equation} 

For any subset $B$ of $E_X^n$ and $\sigma\in \mathfrak S_n$, we denote by $\sigma\cdot B$ the set $\set{\sigma \cdot \V x}{\V x\in B}$.

\begin{IEEEproof}[Proof of Proposition \ref{prop:same_prob_for_given_type}]
The action of $\mathfrak S_n$ is  transitive when restricted to $T_P^{(n)}$. We choose $\sigma$ such that $\sigma\cdot \V y = \V y'$. 

Remark that if  $\V x\in T_{\delta,\delta'}^{(n)}(\V y)$, then $\sigma\cdot x \in T_{\delta,\delta'}^{(n)}(\sigma \cdot \V y)$. In fact, since $(x_1,...,x_n) \mapsto  f(x_1) \cdots f(x_n)$ is invariant under permutations, $\sigma\cdot \V x$ belongs to $ W_{\delta}^{(n)}(\rho)$ too; that it corresponds to $\sigma\cdot \V y$ under $\pi^{\times n}$ is a consequence of \eqref{eq:equivariance}. We conclude that $\sigma(T_{\delta,\delta'}^{(n)}(\V y)) \subset T_{\delta,\delta'}^{(n)}(\V y')$. An analogous argument with $\sigma^{-1}$ gives the equality:
\begin{equation}\label{eq:equality_permuted_set}
\sigma(T_{\delta,\delta'}^{(n)}(\V y)) = T_{\delta,\delta'}^{(n)}(\V y').
\end{equation}

Now, from the definition of the product measure, it is clear that 
\begin{align}
\rho^{\otimes n}(A_1 \times \cdots \times A_n) &= \rho^{\otimes n}(A_{\sigma(1)}\times \cdots \times A_{\sigma(n)})  \\&= \int_{E_X^{n}} 1_{A_1\times \cdots \times A_n}(\sigma^{-1}\cdot \V x) \d \rho(x).
\end{align}
We conclude then that the measures $\rho^{\otimes n}$ and $B\mapsto \int_{E_X^{n}} 1_B(\sigma^{-1}\cdot \V x) \d \rho(x)$ coincide on the $\pi$-system of rectangles $A_1 \times \cdots \times A_n$; by \cite[Lem.~1.6]{Williams1991}, they coincide on the product $\sigma$-algebra $\salg B_X^{n}$; in other words, $\rho$ is invariant under permutations, so \eqref{eq:equality_measure_fibers_type} follows from \eqref{eq:equality_permuted_set}. The same argument works when $\mu$ is finite. 
%The similar proof can be used for $\mu^{\otimes n}$ after decomposing $E_X^n$ into countably many disjoint rectangles of finite measure. 
\end{IEEEproof}

We present this as a first step towards finding a \emph{lower bound} for the volume $\mu^{\otimes n} (T_{\delta,\delta'}^{(n)}(\V y))$ of a doubly typical fiber in terms of the conditional entropy $H(X|Y)$. In fact, such a lower bound would be implied by a lower bound of $\rho^{\otimes n} (T_{\delta,\delta'}^{(n)}(\V y))$, by a standard argument (see the proof of Proposition \ref{prop:upper_bound_vol_T}). Hence one would need to find conditions under which the $(\rho^{\times n})$-probability $p$ of the doubly typical set  $T_{\delta,\delta'}^{(n)}$, lower bounded by $1-2\epsilon$, is divided ``almost equally'' among all the typical fibers; since there are approximately $\e^{n H(Y)}$ of those fibers this could give a lower bound for $\rho^{\otimes n} (T_{\delta,\delta'}^{(n)}(\V y))$ close to $p \e^{-n H(Y)}$. The proposition above shows that there is indeed equidistribution of the probability among fibers of the same type. So the problem is reduced to controlling the possible discrepancy between $\rho^{\otimes n}(T_P^{(n)})$ and $\rho^{\otimes n}(T_{P'}^{(n)})$ when $P\neq P'$; it might be enough to know what happens with adjacent types i.e. when $d_{TV}(P,P')=1/n$. We think that global bounds for the quotients $\diff{\rho_i}{\mu_i}/\diff{\rho_j}{\mu_j}$, for $i\neq k$, might be enough to explicitly control these discrepancies. However, this is a delicate problem that we leave for a future article.

\section{Stratified measures}\label{sec:stratified}

We review first some fundamental facts about rectifiable measures, before introducing the definition of \emph{stratified measure}, a particular case of piecewise continuous measure with a rich geometric structure.

\subsection{Rectifiable measures}\label{sec:rectifiable_measures}

Let $\rho$ be a locally finite measure and $s$ a nonnegative real number. 
Marstrand proved that if the limiting density 
\begin{equation}\label{eq:tangent_density}
\Theta_s(\rho,x) := 
\lim_{r\downarrow 0} \frac{\rho(B(x,r))}{w_s r^s}
\end{equation}
exists and is strictly positive and finite for $\rho$-almost every $x$, then $s$ is an integer not greater than $n$. Later Preiss proved that such a measure is also $s$-rectifiable in the sense of the following definition. For details, see e.g. \cite{DeLellis2008}.

\begin{definition}\label{def:rectifiable_measure} A Radon outer measure $\mu$ on $\Rr^d$ is called $m$-rectifiable if $\mu \ll \mathcal H^m$ and there exists a countably $(\mathcal H^m,m)$-rectifiable Borel set $E$ such that $\mu(\Rr^d\setminus E)=0$ \cite[Def.~16.6]{Mattila1995}.
\end{definition}

Since $\Rr^d$ is countably $d$-rectifiable and $\mathcal H^d = \lambda^d$, an absolutely continuous measure  on $\Rr^d$   is $d$-rectifiable. Similarly, a purely discrete measure is $0$-rectifiable because a countably $0$-rectifiable set is a countable set and $\mathcal H^0$ is the counting measure. 

Remark that for any Borel set $B$ of $\mathcal H^m$, the intersection $B\cap E$ is also Borel, hence $\mathcal H^m$-measurable. Therefore, the restriction of $\mathcal H^m$ to the subspace $\sigma$-algebra $\salg B(E)$ defined in Section \ref{sec:rectifiable_sets} is a measure; this measure is $\sigma$-finite (see Remark \ref{rmk:properties_H_measure}). So we can find a sequence of disjoint Borel sets $(E_i)_{i\in \Nn}\subset \salg B (S)$ of finite $\mathcal H^m$-measure whose union is $E$; by monotonicity,
\begin{equation}
\mathcal H^m(E_i \setminus \bigcup_k f_k(A_k)) \leq \mathcal H^m(E\setminus \bigcup_k f_k(A_k)),
\end{equation}
thus each $E_i$ is $(\mathcal H^m,m)$-rectifiable and Lemma \ref{lem:compact_decomposition} applies to it. We conclude that formula \eqref{eq:rep_rectifiable} also holds for $W=E$.

Consider such a rectifiable measure $\mu$ as per Definition \ref{def:rectifiable_measure}. It holds that $\mu \ll \mathcal H^m\mres E$. If we use the representation \eqref{eq:rep_rectifiable}, we see that $\mu(E_0\cap E)=0$, hence $\mu$ is also absolutely continuous with respect to $H^m\mres E^*$, where $E^*$ is the countably $m$-rectifiable set  $ \bigcup_{i\in \Nn} f_i(K_i)$. One can choose a version of the Radon-Nikodym derivative $g=\diff{\mu}{\mathcal H^m\mres E^*}$ that satisfies $g|_{(E\cap E^*)^c} = 0$ and then introduce, for each $i\in \Nn$, the (necessarily bounded) Borel subset $C_i$ of $K_i$ on which $g>0$. The set $ \tilde E = \bigcup_{i\in\Nn} f_i(C_i)$ is evidently a subset of $E$ and $\mu \ll \mathcal H^m\mres \tilde E$; moreover $g|_{\tilde E}$ serves as a version of the derivative $\diff{\mu}{\mathcal H^m\mres \tilde E}$ and it is positive $(\mathcal H^m\mres E)$-almost everywhere \cite[App.~A]{Koliander2016}. 

By a \emph{carrier} of an $m$-rectifiable measure $\mu$ we mean a countably $m$-rectifiable set $E$ such that $\mu \ll \mathcal H^m\mres E$. Both $E^*$ and $\tilde E$ are carriers.\footnote{Ref. \cite{Koliander2016} uses the term support for $\tilde E$, but this could be confused with the topological notion of support, $$\sup\mu = \Rr^d \setminus \bigcup \set{V}{V\text{ open and } \mu(V)=0}.$$} Their Hausdorff dimension is $m$ by the following lemma.

\begin{lemma}\label{lem:dimension_carrier}
Let $\mu$ be an $m$-rectifiable nonzero measure, and $E$ any countably $(\mathcal H^m, m)$-rectifiable set such that $\mu \ll \mathcal H^m\mres E$. Then the Hausdorff dimension of $E$ is $m$.  
\end{lemma}
\begin{IEEEproof}
Write $E$ as in \eqref{eq:rep_rectifiable}. Remark that $0\leq  \mathcal H^m(f_i(K_i)) < \infty$ for every $i\in\Nn$; this means that the Hausdorff dimension of each set $f_i(K_i)$ is at most $m$. Since the $m$-Hausdorff measure of $E$ cannot vanish (otherwise $\mu$ would be zero), we conclude that for some $j\in \Nn$, $\mathcal H^m(f_j(K_j))>0$, therefore $\dim_H f_j(K_j)=m$. The result follows from he countable stability of the Hausdorff dimension: if $F_0,\, F_1,\, F_2,\,...$ is a countable sequence of sets, then $\dim_H \bigcup_{i\in \Nn} F_i = \sup_{i\in \Nn} \dim_H F_i$ \cite[p. 49]{Falconer1990}.
\end{IEEEproof}

\begin{lemma}\label{lemma:products_carriers}
If $S_1$ is a carrier of Hausdorff dimension $m_1$ and $S_2$ a carrier of Hausdorff dimension $m_2$, then $S_1\times S_2$ is also a carrier, of Hausdorff dimension $m_1+m_2$. 
Additionally, the Hausdorff measure $\mathcal H^{m_1+m_2}|_{S_1\times S_2}$ equals $\mathcal H^{m_1}|_{S_1} \otimes \mathcal H^{m_2}|_{S^2}$. 
\end{lemma}
\begin{IEEEproof}
For $i\in \{1,2\}$, consider representations $S_i=\bigcup_{k\in \Nn} f^i_k(C^i_k)$ for certain bounded Borel sets $\{C^i_k\}_k$ of $\Rr^{m_i}$ and Lipschitz functions $\{f_k:\Rr^{m_i}\to\Rr^d\}_k$ such that each $f_k$ is injective on $C_k$. In this case, $S_1\times S_2$ is countably $(m_1+m_2)$-rectifiable, for it can be written as $\bigcup_{k,l} (f^1_k \times f^2_l)(C^1_k\times C^2_l)$. Moreover, one can prove that 
\begin{equation}\label{eq:product-hausdorff}
\mathcal H^{m_1+m_2}|_{S_1\times S_2} = \mathcal H^{m_1}|_{S_1} \times H^{m_2}|_{S_2}
\end{equation}
as follows \cite[Lem.~27]{Koliander2016}: each set $f^i_k(C^i_k)$ is Borel (see footnote \ref{footnote_Lusin_souslin}), therefore $\mathcal H^{m_1+m_2}|_{f_k^1(C_k^1)\times f_l^2(C_l^2)} = \mathcal H^{m_1}|_{f_k^1(C_k^1)} \times H^{m_2}|_{f_l^2(C_l^2)}$ \cite[Thm.~3.2.23]{Federer1969}; this simply comes from the diagonal form of the corresponding Jacobian matrix that appears in the \emph{area formula}. The more general formula \eqref{eq:product-hausdorff} is a consequence of $\sigma$-additivity.
\end{IEEEproof}

It is worth remarking that, in general, the product of an $(\mathcal H^{m_1},m_1)$-rectifiable set and an $(\mathcal H^{m_2},m_2)$ rectifiable set is \emph{not} $(\mathcal H^{m_1+m_2},m_1+m_2)$-rectifiable, see \cite[3.2.24]{Federer1969}. This potential problem is overcome here by the restriction of $\mu$ to its carrier, which accepts an explicit representation as a countable union of disjoint Borel sets. The possibly pathological $S_0$ becomes irrelevant if we only care about the measure $\mu$. 

\subsection{Definition of stratified measures}
\label{sec:stratified_def}

\begin{definition}[$k$-stratified measure]
A measure $\nu$ on $(\Rr^d,\mathcal B(\Rr^d))$ is \emph{$k$-stratified}, for $k\in \Nn^*$,  if there are integers $(m_i)_{i=1}^k$ such that  $0 \leq m_1 < m_2 <... <m_k \leq d$ and $\nu$ can be expressed as a sum $\sum_{i=1}^k \nu_i$, where each $\nu_i$ is a nonzero $m_i$-rectifiable measure.
\end{definition}

Thus  $1$-stratified measures are rectifiable measures. If $\nu$ is $k$-stratified for some $k$ we simply say that $\nu$ is a \emph{stratified measure.}

As we discussed in \ref{sec:rectifiable_measures}, a nonzero measure $\nu$ on $\Rr^d$ is $m$-rectifiable if there exists a  $\mathcal H^m$-measurable, countably $(\mathcal H^m,m)$-rectifiable set $S$ \emph{of positive $\mathcal H^m$-measure} such that $\nu\ll\mathcal H^m\mres S$. Moreover,  $\nu$ coincides with its restriction $\nu \mres {\tilde S}$ to a subset $\tilde S$ of $S$ of the form $\bigcup_{i\in \Nn} f_i(C_i)$, called the \emph{carrier}, such that, for each $i\in \Nn$, $f_i:\Rr^m\to \Rr^d$ is a Lipschitz function, $C_i$ a bounded Borel set, and $f_i|_{C_i}$ an injective map.  Each $f_i(C_i)$ is Borel (see footnote \ref{footnote_Lusin_souslin}). The Hausdorff dimensions of $S$ and $S'$ equal $m$ by Lemma \ref{lem:dimension_carrier}. 

Hence for $\nu_i$ in Definition \ref{def:rectifiable_measure}, we can introduce a carrier $E_i= \bigcup_{k\in \Nn} f^i_k(C_k)$, where the $C^i_k$ are bounded Borel subsets of $\Rr^{m_i}$ and each $f^i_k:\Rr^{m_i}\to \Rr^d$ is a  Lipschitz function injective on $C^i_k$. Remark that $\mathcal H^m(E_i)=0$ for any $m>m_i$. This implies in particular that, for any $i\geq 2$, the measure $\nu_i$ coincides with the restricted measure $\nu_i \mres E_i'$, where $E_i':=  E_i\setminus \bigcup_{j=0}^{i-1}E_j$,  and evidently $\mu \ll \mathcal H^m \mres {E_i'}$; moreover, $E_i'$ can be written as $\bigcup_{k\in \mathbb N} f^i_k(C_k\cap \bigcup_{j=0}^{i-1}(f_k^i)^{-1}(E_j^c))$ and clearly $C_k\cap \bigcup_{j=0}^{i-1} (f_k^i)^{-1}(E_j^c)$ is  also a bounded Borel set. Therefore, \emph{there is no loss of generality in supposing that the carriers $\{E_i\}_{i=1}^k$ are disjoint, as we did in the introduction.} 

\begin{example}
The  decomposition of a measure $\nu$ on $(\Rr^d,\mathcal B(\Rr^d))$ as a sum $\nu_{a.c.} + \nu_{p.d.}$, where $\nu_{a.c.}$ is absolutely continuous with respect to the Lebesgue measure on the $d$-rectifiable set $E_2=\Rr^d$ and $\nu_{p.d.}$ is supported on a $0$-rectifiable set $E_1$ exhibits $\nu$ as a stratified measure. We see that we can restrict $\nu_{a.c.}$ to $E_2\setminus E_1$ without changing $\nu_i$ or $\nu$ as measures on $(\Rr^d,\mathcal B(\Rr^d))$.

For instance, the measure $\rho = \frac 12 \cf_{[0,1]}\lambda^{d} + \frac 12 \delta_{1/2}$, where $\lambda^{d}$ is the Lebesgue measure, equals  $\rho = \frac 12 \cf_{[0,1]\sm\{1/2\}}\lambda^{d} + \frac 12 \delta_{1/2}$.
\end{example}

\begin{example} Another example is $\nu = \sum_{k=1}^n \nu_i$, where $\nu_i$ is supported on an $m_i$-dimensional $C^1$-submanifold $E_i$ of $\Rr^d$, see Example \ref{ex:manifolds_rectifiable}. The construction  above exhibits $\nu$  as a sum of measures $\sum_{i=1}^k \tilde \nu_i$, each carried by a rectifiable set $ E_i'$, such that the sets $\{ E_i'\}_{i=1}^k$ are pairwise disjoint. Remark however that these sets are not manifolds anymore, which gives another reason to work with rectifiable sets instead. 
\end{example}

\begin{remark}
A more general  definition would be: a measure $\nu$  on a \emph{metric space} $(E_X,d_X)$ can be decomposed as a sum of measures $\nu_1,...,\nu_k$, and each $\nu_i$ is an $m_i$-rectifiable measure. Extensions of our results to this case appear as a natural problem. As far as we know, Lemma \ref{lem:compact_decomposition} and the representation \eqref{eq:rep_rectifiable} are not directly available (with injective $f_k$).  
\end{remark}

\begin{definition}[Standard form]\label{def:stratified_standard}
Given a stratified \emph{probability} measure $\rho$, we shall write it, without loss of generality, in the \emph{standard form} 
\begin{equation}\label{eq:stratified_proba_reduced}
\rho = \sum_{i=1}^k q_i \rho_i
\end{equation}
where each $\rho_i$ is a rectifiable probability measure with carrier $E_i$ of dimension $m_i$ (so that $\rho_i = \rho_i\mres E_i$), the carriers $(E_i)_{i=1}^k$ are disjoint, $0 \leq m_1 < \cdots < m_k \leq d$, and $(q_1,...,q_k)$ is a probability vector \emph{with strictly positive entries.}
\end{definition}

\begin{remark}
Define  $E_X = \bigcup_{i=1}^k E_i$. Since $\rho = \rho\mres E_X$, we suppose---with no loss of generality---that $\rho$ is defined on $(E_X, \salg B(E_X))$. Following the notations of Section \ref{sec:piecewise_continuous}, we introduce the set $E_Y=\{1,...,k\}$ and the projection $\pi:E_X\to E_Y$ that maps $E_i$ to $i$. For each measure $\rho_i$ is absolutely continuous with respect to the measure $\mu_i := \mathcal H^{m_i}\mres E_i$, equation \eqref{eq:stratified_proba_reduced} presents $\rho$ as a piecewise continuous probability measure in mixture form, see Section \ref{sec:chain_rule}.
\end{remark}

\subsection{Products of rectifiable measures}
\label{sec:products_rectifiable_measures}

The product $\rho^{\otimes n}$ of $n$-copies of $\rho$ satisfies
\begin{equation}
\rho^{\otimes n} = \sum_{\V y=(y_1,...,y_n) \in E_Y^n}  q_1^{N(1;\V y)} \cdots q_k^{N(k;\V y)} \rho_{y_1}\otimes \cdots \otimes \rho_{y_n}. 
\end{equation}
By standard properties of product measures, each measure $\rho_{\V y} := \rho_{y_1}\otimes \cdots \otimes \rho_{y_n}$ is absolutely continuous with respect to $\mu_{\V y}:= \mu_{y_1} \otimes \cdots \otimes \mu_{y_n}$. For any $\V y\in E_Y^n$, define $m(\V y) := \sum_{j=1}^n \dim E_{y_j}$. Lemma \ref{lemma:products_carriers} and its proof  show that  the stratum $\Sigma_{\V y} := E_{y_1}\times \cdots \times E_{y_n}$  can be written as a countable union of images of bounded Borel subsets of $\Rr^{m(\V y)}$ under Lipschitz maps, hence that $\Sigma_{\V y}$ is countably $m(\V y)$-rectifiable set. Moreover, Lemma \ref{lemma:products_carriers} also shows  that the product measure $\mu_{\V y}$ equals $\mathcal H^{m(\V y)}\mres \Sigma_{\V y}$. Therefore each measure $\rho_{\V y}$ is rectifiable. 

\begin{proposition}
Let $\rho$ be a stratified measure. Then the product measures $\rho^{\otimes n}$ are stratified too, for every $n\geq 1$. 
\end{proposition}
\begin{IEEEproof}
To exhibit $\rho^{\otimes n}$ as a stratified measure, one must simply group these measures according to their dimension: first define the set 
\begin{equation}
M = \bigset{ m(\V y) }{\V y \in E_Y^n},
\end{equation}
and then write
\begin{equation}
\rho^{\otimes n} = \sum_{m\in M}\left(  \sum_{\V y\::\; m(\V y) = m} q_1^{N(1;\V y)} \cdots q_k^{N(k;\V y)} \rho_{\V y}\right).
\end{equation}
For  each $M\in M$,  $\sum_{\V y\::\; m(\V y) = m} q_1^{N(1;\V y)} \cdots q_k^{N(k;\V y)} \rho_{\V y}$ is a rectifiable measure, being absolutely continuous with respect to $\mathcal H^m$ restricted to $\bigcup_{\V y\::\; m(\V y) = m} \Sigma_{\V y}$.
\end{IEEEproof}

\subsection{Typical strata}
\label{sec:typical_strata}

 The doubly typical strata $T^{(n)}_{\delta, \delta'}(\V y)$ is a subset of $ \Sigma_{\V y} = (\pi^{\times n})^{-1}(\V y)$, therefore $\mu^{\otimes n}(T^{(n)}_{\delta, \delta'}(\V y)) = \mu_{\V y}(T^{(n)}_{\delta, \delta'}(\V y)) = \mathcal H^{m(\V y)}(T^{(n)}_{\delta, \delta'}(\V y)).$ It follows from Propositions \ref{prop:upper_bound_vol_T} and \ref{prop:tightness_bound} that the conditional entropy $H(X|Y)$ gives an asymptotically tight upper bounds for the exponential rate of growth, in $n$, of the \emph{$m(\V y)$-Hausdorff measure} of the doubly typical fiber $T^{(n)}_{\delta, \delta'}(\V y)$, for each $\V y\in A_{\delta'}^{(n)}$. 
 
 The measure $\rho^{\otimes n}$ concentrates asymptotically on the doubly typical strata. For any $\delta, \, \delta' > 0$, define 
\begin{equation}\label{eq:restriction}
\rho^{(n)} := \rho^{\otimes n}|_{T^{(n)}_{\delta,\delta'}} = \sum_{ \V y \in A_{\delta'}^{(n)}} \rho^{\otimes n}|_{T^{(n)}_{\delta,\delta'}(\V y)}.
\end{equation}

\begin{proposition}\label{prop:approximation_product_measure}
For all $\delta, \delta',\epsilon>0$, there exists $n_0\in \Nn$ such that if $n\geq n_0$, then $d_{TV}(\rho^{\otimes n}, \rho^{(n)}) <\epsilon$. 
\end{proposition}
\begin{IEEEproof}
Recall that for  measures $\xi, \, \xi'$ on $\salg B(E_X)$, the total variation distance $d_{TV}$ is given by
\begin{equation}
d(\xi,\xi'):=\sup_{A\in \salg B(E_X)} |\xi(A) -\xi'(A)|
\end{equation} 
Let $\epsilon > 0$ be given.  For any measurable $A$, 
\begin{equation}
\rho^{\otimes n}(A) -\rho^{(n)}(A) = \int_{E_X} \cf_{A\cap (T^{(n)}_{\delta,\delta'})^c} \d\rho.
\end{equation}
 By monotonicity of the integral, this expression is maximized when $A=(T^{(n)}_{\delta,\delta'})^c$, and takes the value $\rho^{\otimes n}((T^{(n)}_{\delta,\delta'})^c)$, which by the union bound is smaller or equal than $\rho^{\otimes n} ((W_{\delta}^{(n)}(\rho))^c) + (\pi_*\rho)^{\otimes n}((A_{\delta'}^{(n)})^c).$ The first term can be made smaller than $\epsilon/2$ for $n\geq n_1$, for certain $n_1\in \Nn$, according to the AEP (Proposition \ref{prop:AEP}). The same holds for the second term, according to our initial definition of $A_{\delta'}^{(n)}$ in Section \ref{sec:cond_entrop_asymp}. 
\end{IEEEproof}

\begin{remark}
The same proof would give nonasymptotic concentration inequalities provided one can control explicitly $\rho^{\otimes n} ((W_{\delta}^{(n)}(\rho))^c)$ and $(\pi_*\rho)^{\otimes n}((A_{\delta'}^{(n)})^c)$. Chebyshev's inequality gives an upper bound for $\rho^{\otimes n} ((W_{\delta}^{(n)}(\rho))^c)$ when $-\ln \diff\rho\mu$ has finite variance; one might also use Hoeffding's inequality if $|-\ln \diff\rho\mu|$ is bounded. In turn, if we work with $A_{\delta'_n}^{(n)}$ given by the strongly typical sequences of Section \ref{sec:strong_typicality}, then  $(\pi_*\rho)^{\otimes n}((A_{\delta_n'}^{(n)})^c) \leq \epsilon_n = 2|E_Y|\e^{-2n\eta_n^2}$; one may take $\eta_n = O(n^{1/2-\xi})$, for any $\xi>0$.
\end{remark}

\subsection{Typical dimensions and asymptotic equipartition}\label{sec:typical_dimension}

Let us start with an informal description of our argument. Consider a stratified probability measure in the standard form \eqref{eq:stratified_proba_reduced}, and the projection $\pi$ just introduced from $E_X = \bigcup_{i=1}^kE_i$ to $E_Y=\{1,...,k\}$. Remark that for every $n\in \Nn$, the probability $\pi^{\times n}_*(\rho^{\otimes n})$ equals $(\pi_*\rho)^{\otimes n}$, so we may regard $\V y \in E_Y^n$ as the realization of an i.i.d. sequence $\V Y_n =(Y_1,...,Y_n)$ of $E_Y$-valued random variables, each one with law $\pi_*\rho$. We denote by $Q$ the p.m.f. of this law, which maps $y_i$ to $q_i$.  As $n\to\infty$, the measure $(\pi_*\rho)^{\otimes n}$ concentrates around certain typical sequences $\V y \in E_Y^n$, such that $y_i$ appears about $q_in$  times in $\V y$. This means that in the corresponding set $E_n(\V y)$, the factor $E_i$ also appears  about $q_in$ times, so
\begin{equation}
\dim(E_n(\V y)) \approx \sum_{i=1}^k (q_in)m_i
\end{equation}
is the ``typical  dimension'' of a stratum $E_n(\V y)$. Moreover, only ``typical strata'' that correspond to typical sequences $\V y\in E_Y^n$ are asymptotically relevant, and their dimensions are close to the typical one. 

In fact, the Hausdorff dimension of a stratum is  a random variable $D_n := m(\V Y_n) = \sum_{i=1}^n m_{Y_i}$ with mean $\mathbb E(D_n) = n \mathbb E(D_1)= n \sum_{j=1}^k q_j m_j$, and the variable $D_n$ concentrates asymptotically around its mean. 

More formally, we introduce the set $A_{\delta_n'}^{(n)}$ of strongly $(Q,\eta_n)$-typical sequences, with the $\eta_n=n^{-1/2+\xi})$ chosen according to the $\eta$-convention introduced at the end of Section \ref{sec:typical_dimension} and $ \delta'_n=-|E_Y|\eta_n\ln\eta_n$.  We have mentioned that, in this case,  
\begin{equation}
(\pi_*\rho)^{\otimes n}(A_{\delta_n'}^{(n)})\geq 1-\epsilon_n.
\end{equation}
On this set of high probability, we have the following estimate for the dimension $D_n$. 
\begin{proposition}\label{prop:typical-dimensions}
For any $\V y\in A_{\delta_n'}^{(n)}$, 
\begin{equation}\label{eq:interval_typical_dimensions}
m(\V y) \in [n \mathbb E(D_1) - n^{1/2+\xi}, n\mathbb E(D_1) + n^{1/2+\xi}].
\end{equation}
\end{proposition}
\begin{IEEEproof}
By definition of strong typicality, for any sequence $\V y\in E_Y^n$ with associated empirical measure $\tau_{\V y}$ with p.m.f. $P$, it holds that $P\ll Q$ and
\begin{equation}
\forall a\in E_Y, \quad |P(a) - Q(a)| \leq \eta_n.
\end{equation}
In other words, for any $a\in E_Y$, we have the inequalities
\begin{equation}
n(Q(a) - \eta_n) \leq N(a;\V y) \leq n(Q(a) +\eta_n). 
\end{equation}
This entails
\begin{equation}\label{eq:inequalities_appearances}
nQ(a) - n^{1/2+\xi} \leq N(a;\V y) \leq nQ(a) +n^{1/2+\xi},
\end{equation}
for a given $\xi>0$. The proposition follows  from \eqref{eq:inequalities_appearances} and the formula  $m(\V y) =\sum_{i=1}^k N(i;\V y) m_i$.
\end{IEEEproof}

At this point we can put all the pieces together to state a full asymptotic equipartition theorem for stratified measures. 

\begin{theorem}
Let $\rho$ be a stratified measure as in Definition \ref{def:stratified_standard} and let $\delta>0$ be given. Fix a sequence $(\eta_n)_n  $ such that $\eta_n=n^{-1/2+\xi}$ and set  $\delta'_n=-|E_Y|\eta_n\ln\eta_n$. Let $A_{\delta'_n}^{(n)}$ be the set of $(Q,\eta_n)$-typical sequences in $E_Y^n$, $T^{(n)}_{\delta,\delta'_n}=(\pi^{\times n})^{-1}(A_{\delta'_n}^{(n)})\cap W_{\delta}^{(n)}(\rho)$ be the set of  doubly  typical sequences in $E_X^n$, and $T^{(n)}_{\delta,\delta'_n}(\V y) = T^{(n)}_{\delta,\delta'_n}\cap (\pi^{\times n})^{-1}(\V y)$ a doubly typical stratum for any $\V y\in A_{\delta'_n}^{(n)}$. For any $\epsilon > 0$ there exists an $n_0\in \Nn$ such that for any $n\geq n_0$ the restriction  of $\rho$ to $T^{(n)}_{\delta,\delta'_n}$, 
 $$\rho^{(n)}=\sum_{ \V y \in A_{\delta'}^{(n)}} \rho^{\otimes n}|_{T^{(n)}_{\delta,\delta'}(\V y)},$$ satisfies $d_{TV}(\rho^{\otimes n},\rho^{(n)})<\epsilon$.
  Moreover, the measure $\rho^{(n)}$ is stratified and can be written as a sum of $m$-rectifiable measures for $m$ in $[n \mathbb E(D_1) - n^{1/2+\xi}, n\mathbb E(D_1) + n^{1/2+\xi}]$. The conditional entropy $H(X|Y)=\sum_{i=1}^k q_i \ent{\mu_i}(\rho_i)$ quantifies the volume growth of most doubly typical fibers, in the following sense: 
 \begin{enumerate}
 \item   For any $\V y \in A_{\delta'}^{(n)}$,
 $$\frac{1}{n} \ln \mathcal H^{m(y)}(T^{(n)}_{\delta,\delta'}(\V y)) \leq H(X|Y) + (\delta + \delta'_n),$$
 where $m(\V y) =\sum_{i=1}^n m_{y_i}$.
 \item For any $\epsilon > 0$, the set $B_{\epsilon}^{(n)}$  of $\V y\in \subset A_{\delta'_n}^{(n)}$ such that 
 $$\frac{1}{n} \ln \mathcal H^{m(y)}(T^{(n)}_{\delta,\delta'}(\V y)) > H(X|Y) -\epsilon + (\delta + \delta'_n),$$
 satisfies 
 $$\limsup_{||(\delta,\delta'_n)||\to 0}\limsup_{n\to \infty} \frac{1}{n} \ln |B^{(n)}_\epsilon| = H(Y).$$
 \end{enumerate}
\end{theorem}
\begin{IEEEproof}
The first claim is Proposition \ref{prop:approximation_product_measure}. For the next one, just remark that $\rho^{\otimes n}|_{T^{(n)}_{\delta,\delta'}(\V y)}$ is absolutely continuous with respect to $\mathcal H^{m(\V y)}\mres \Sigma_{\V y}$ hence rectifiable; the typical dimensions are concentrated around $\mathbb E(D_1)$ according to Proposition \ref{prop:typical-dimensions}. The volume estimates in terms of $H(X|Y)$ are restatements of Propositions \ref{prop:upper_bound_vol_T} and \ref{prop:tightness_bound}; just remark that $\mu^{\otimes n}|_{\Sigma_{\V y}} = \mu_{y_1} \times \cdots \mu_{y_n} = \mathcal H^{m(\V y)}$ as we explained at the end of Section \ref{sec:stratified}.
\end{IEEEproof}

\section{Connection with R\'enyi's dimension and entropy}
\label{sec:Renyi-dim-and-ent}

The theory developed in the previous sections gives a very concrete probabilistic meaning to the mean dimension $\mathbb E(D_1)=\sum_{i=1}^k q_i m_i$ of the stratified measure \eqref{eq:stratified_proba_reduced} as well as its generalized entropy $H_{\mu}(\rho)$. In this section, we study the relation between these quantities and the  dimension and entropy introduced by R\'enyi in \cite{Renyi1959}. 

Consider a random vector $\V X =(X_1,...,X_d) \in \Rr^d$. We denote by $[\V X]$ the vector in $\Rr^d$ whose $i$-th component is the integer part $[X_i]$ of $X_i$, and set $\V X_n = [n \V X]/n$ for any $n\geq 1$. 

R\'enyi remarked that when the law 
 $\rho$ of $\V X$ has a density with respect to the $d$-dimensional Lebesgue measure $\lambda^d$ and both the differential entropy $H_{\lambda^d}(\rho)$ and the discretized entropy $\ent\#([\V X]) $ are finite, the asymptotic formula
\begin{equation}\label{eq:asymptotic_expansion_discretization}
\ent\#(\V X_n) = d\ln n + \ent{\lambda^d}(\rho) + o(1)
\end{equation}
holds (see \cite[Thm.~4]{Renyi1959}) and he took this as a starting point to generalize the notion of dimension as follows.\footnote{Asymptotic expansions similar to \eqref{eq:asymptotic_expansion_discretization} appeared also in the article ``Amount of information and entropy for continuous distributions'', published by Kolmogorov in 1958 \cite[p.~33ff]{Kolmogorov1993}; they are also the subject of an article published by Csisz\'ar in 1973 \cite{Csiszar1973}, which generalizes \cite[Thm.~4]{Renyi1959}.}

\begin{definition}[Information dimension]
The information dimension of the random variable $\V X$ is 
\begin{equation}
\dim_I X := \lim_n \ent\#(\V X_n)/\ln n,
\end{equation}
provided the limit exists.
\end{definition}

This definition only depends on the law $\rho$, hence we also write $\dim_I \rho$ instead of $\dim_I X$. More explicitly, for each $\V k=(k_1,...,k_d) \in \Zz^d$, we introduce the cube $C_n(\V k) = [\frac {k_1}{n}, \frac{k_1 + 1}{n})\times \cdots \times   [\frac {k_d}{n}, \frac{k_d + 1}{n})$; these cubes form a partition of $\Rr^d$, associated with the mesh $\Zz^d/n$. For each $n\geq 1$ and $\V k \in \Zz^d$, we  define the probabilities $p_{n,\V k} := \rho(C_n(\V k))$, which satisfy
\begin{align}
p_{n,\V k} 
& = \mathbb P\left( \frac{k_i}{n} \leq X_i < \frac{k_i+1}{n}, \, i\in \{1,...,d\}\right)\\
&=\mathbb P(\V X_n = \V k/n).
\end{align}
Therefore $\ent\#(\V X_n) = -\sum_{\V k\in \Zz^d} p_{n,\V k} \ln p_{n,\V k}.$ 

Elaborating on previous work in dynamical systems, particularly \cite{Young1982} and \cite{Hunt1997}, Verd\'u and Wu \cite[App. I]{Wu2010} proved that
\begin{equation}\label{eq:Verdu-Wu}
\dim_I \rho = \lim_{\epsilon \to 0} \mathbb E_{\rho} \left( \frac{ \ln \rho (B(X,\epsilon))}{\ln\epsilon} \right),
\end{equation}
where $B$ denotes an Euclidean ball. This can be used to compute the information dimension of a rectifiable measure as follows. In virtue of Preiss' theorem, a measure  $\rho$ is  $m$-rectifiable  if and only if the density $\Theta_m(\rho,x)$ in \eqref{eq:tangent_density} exists and is finite and nonzero for $\rho$-almost every $x$. It is easy to see that this implies 
\begin{equation}
\lim_{\epsilon \to 0} \frac{\ln \rho (B(X,\epsilon))}{\ln\epsilon} = m \quad \rho\text{-a.e.}
\end{equation}
Therefore $\dim_I\rho = m$, provided one can exchange the limit and the expectation in \eqref{eq:Verdu-Wu} e.g. if $\ln \rho B_2(X,\epsilon)/\ln\epsilon$ is dominated by an integrable function for $\epsilon$ sufficiently small. Ignoring this condition leads to contradictions: \cite[p. 197-198]{Renyi1959} introduces a continuous variable $\V X$ on $\Rr^1$ with density function $f(x)=\sum_{n=2}^{\infty} C\ln^{-2}(n)\cf_{[n,n+1/n]}$, where $C$ is a normalization factor; although the law of $\V X$ is   $1$-rectifiable its information dimension diverges because  $H_{\#}(\V X_1)$ is infinite and $H_{\#}(\V X_1) \leq H_{\#}(\V X_n)$ for any $n\geq 1$ as a consequence of the Data Processing Inequality.

We say that an $m$-rectifiable measure $\rho$ is \emph{dimensional regular} if $\dim_I\rho = m$. Based on R\'enyi's results, we conjecture that $H_{\#}(\V X_1)<\infty$ is a sufficient condition for dimensional regularity

In the case of stratified measures, we have the following result.

\begin{proposition}
Let $\V X$ be an $\Rr^d$-valued random variable with law $\rho$. Suppose $\rho$ is a stratified probability measure in standard form as in Definition \ref{def:stratified_standard}, and that each $\rho$ is a dimensional-regular $m_i$-rectifiable measure. Then
\begin{equation}
\dim_I \V X =\sum_{i=1}^k q_i m_i.
\end{equation}
\end{proposition}
This   proposition follows from \cite[Thm. 2]{Wu2010} (see also \cite[Lem. 4]{Charusaie2022}). We provide here an alternative proof, inspired by \cite{Renyi1959}.
\begin{IEEEproof}
Set $p_{n,\V j} = \rho(C_{n,\V j})$ and $p^i_{n,\V j} = \rho_i(C_{n,\V j})$. By definition, 
\begin{equation}
p_{n,\V j} = \sum_{i=1}^k q_i p_{n,\V j}^i.
\end{equation}

The fundamental ingredient of our proof is the log-sum inequality \cite[Lem. 3.1]{Csiszar1981}: for arbitrary nonnegative numbers $(a_i)_{i=1}^n$ and $(b_i)_{i=1}^n$, we have
\begin{equation}
\sum_{i=1}^n a_i \ln \frac{a_i}{b_i} \geq  a \ln \frac{a}{b},
\end{equation}
where $a=\sum_{i=1}^n a_i$ and $b=\sum_{i=1}^n b_i$. 

Specialized to the case $b_i=1$, $i=1,...,n$, the inequality can be restated as
\begin{equation}\label{eq:specialized-log-sum}
0\leq a\left(\sum_{i=1}^n \frac{a_i}{a}\ln \frac{a_i}{a} + \ln n\right),
\end{equation}
whose right-hand side is not greater than $a\ln n$. 

Therefore, if we set  $n=k$ and  $a_i = p^i_{n,\V j}$ for any $i=1,...,k$, we can conclude from \eqref{eq:specialized-log-sum} that
\begin{multline}
- p_{n,\V j} \ln k \leq  \sum_{i=1}^k q_i p_{n,\V j}^i \ln (q_ip_{n,\V j}^i) - p_{n,\V j} \ln p_{n,\V j} 0  \leq 0,
\end{multline}
and then, summing over $\V j\in \Zz^d$, 
\begin{equation}
-\ln k \leq - \left(H(Y) + \sum_{i=1}^k q_i H_{\#}(\V X^{(i)}_n)\right) + H_{\#}(\V X_n)  \leq 0, 
\end{equation}
where $\V X^{(i)}$ denotes an $\Rr^d$-valued random variable with law $\rho_i$. We divide by $\ln n$ and let $n\to \infty$ to finish the proof.
\end{IEEEproof}

We have seen that our notion of dimension of a stratified measure coincides with R\'enyi's. However, this is not the case for the entropy: if $\rho$ is stratified as in Definition \ref{def:stratified_standard} and $\mu=\sum_{i=1}^{k} \mathcal H^{m_i}\mres E_i$,
\begin{equation}
\ent\mu(\rho)\neq \lim_n \left(\ent\#(\V X_n) - (\dim_I \V X) \ln n \right).
\end{equation}
This discrepancy already happens when $k=1$. The following result makes this explicit. It is similar to Lemma 52 in \cite{Koliander2016}, which a theorem by Csisz\'ar \cite[Thm. 1]{Csiszar1973} that has an involved proof relying on the  notion of \emph{well-approximating partition}. We present here a much simpler and direct martingale argument, valid in the more restricted setting of a sequence of partitions that is totally ordered in the sense of refinement. It might be similar to a proof by Perez mentioned in \cite[p. 4]{Csiszar1973} that we were not able to access.

\begin{theorem}\label{thm:limit_formula_entropy}
\begin{enumerate}
\item \label{thm:limit-formula:part1} Let $\mu$ be a $\sigma$-finite measure and $\rho$ a probability measure absolutely continuous with respect to $\mu$. Then,
\begin{equation}
\diff{\rho}{\mu} = \lim_n \sum_{\substack{\V j \in \Zz^d\\ \mu(C_{2^n,\V j})>0}} \frac{\rho(C_{2^n,\V j})}{\mu(C_{2^n,\V j})}\cf_{C_{2^n,\V j}}
\end{equation}
almost surely and in $L^1$. 
\item \label{thm:limit-formula:part2}  Let $\V X$ be an $\Rr^d$-valued random variable with rectifiable law $\rho$, so that $\rho \ll \mathcal H^m \mres E$ for some $(\mathcal H^m, m)$-rectifiable Borel set $E$. Suppose $\mu:= \mathcal H^m\mres E$ is a finite measure and that $\diff \rho\mu$ is bounded. Then
\begin{equation}\label{eq:limit-entropy-partition}
\ent\mu(\rho) = \lim_n \left( \ent\#(\V X_{2^n}) + \sum_{\substack{\V j\in \Zz^d\\p_{2^n,\V j}>0}} p_{2^n,\V j} \ln \mu(C_{2^n,\V j})\right).
\end{equation}
\end{enumerate}
\end{theorem}

\begin{remark} Equation \eqref{eq:limit-entropy-partition} would also hold under certain conditions that make the entropy continuous as a function of the densities equipped with the $L^1$-norm, like those in \cite{Ghourchian2017}. 
\end{remark}

The term 
\begin{equation}\label{eq:defect_term}
\sum_{\substack{\V j\in \Zz^d\\p_{2^n,\V j}>0}} p_{2^n,\V j} \ln \mathcal H^m(C_{2^n,\V j}\cap E)
\end{equation}
 depends nontrivially on the geometry of $E$.  For instance, if $d=2$ and $E$ is a portion of the line $\{y=x\}$, then $\mathcal H^1(C_{n,\V j} \cap E) = \sqrt{2}/n$ provided $\V j$ and $\V j + (1,1,...,1)$ lie on the segment; more generally, if $E=\{y=mx\}$, then $\mathcal H^1(C_{n,\V j} \cap E) = \sqrt{1+m^2}|\operatorname{proj}_1(C_{n,\V j} \cap E)|$, where the last factor is the  Lebesgue measure of the projection to the $x$-axis (see the area formula in \cite[Thm.~2.71]{Ambrosio2000}). More generally, if $\mathcal H^m(C_{n,\V j}\cap E)= K(\V j,n) (\frac 1n)^m$ for some constant $K(\V j,n)\geq 0$, then  \eqref{eq:defect_term} equals
\begin{equation}
\sum_{\substack{\V j\in \Zz^d\\p_{2^n,\V j}>0}} p_{2^n,\V j} \ln K(\V j, 2^n) - mn\ln 2.
\end{equation}
The expectation $\sum_{\V j\in \Zz^d\,:\,p_{2^n,\V j}>0} p_{2^n,\V j} \ln K(\V j, 2^n)$ can be arbitrarily intricate: consider, for example, a general piecewise linear 1-dimensional submanifold of $\Rr^2$. 

That said, the situation simplifies in the extremal cases considered by R\'enyi in \cite{Renyi1959}. On the one hand, if $E=\Rr^d$, then $\mathcal H^d(C_{n,j}\cap E) = (1/n)^d$ and
\begin{equation}
\ent{\lambda^d}(\rho) = \lim_n \left( \ent\#(\V X_{2^n}) -d\ln (2^n)\right).
\end{equation}
On the other hand, if $E$ is a countable subset of $\Rr^d$ without accumulation points, then eventually $\mathcal H^0(C_{n,\V j}\cap E)$ equals zero or one, hence \eqref{eq:defect_term} vanishes and 
\begin{equation}
\ent{\mathcal H^0}(\rho) = \lim_n \ent\#(\V X_{2^n}) .
\end{equation}

\section{Final remarks}

We list here some problems that remain open. 

Firstly, the determination of  a lower bound for the volume of doubly typical strata $\mu^{\otimes n} (T_{\delta,\delta'}^{(n)}(\V y))$  in terms of the conditional entropy $H(X|Y)$. We outlined an strategy to tackle this problem at the end of Section \ref{sec:symmetry}.

Secondly, we would like to find simple sufficient conditions for the dimensional regularity of a rectifiable measure. Is it enough to suppose that $H_{\#}(\V X_1)<\infty$ (as it is the case in the simpler situations considered by R\'enyi)? This would mean that any $m$-rectifiable measure of finite information dimension satisfies $\dim_I \rho = m$. Young's results \cite{Young1982} are probably relevant, but some of them need to be adapted to  noncompact measure metric spaces. 

Finally, we wonder if there is a simple formula for 
\begin{equation}
\lim_n \sum_{\substack{\V j\in \Zz^d\\p_{n,\V j}>0}} p_{n,\V j} \ln K(\V j, n)
\end{equation}
 in terms of $\rho$ and the Jacobian determinants of the Lipschitz appearing in the representation \eqref{eq:rep_rectifiable}. This would give an explicit quantification of the difference between the generalized entropy $H_{\mu}(\rho)$  and R\'enyi's dimensional entropy.

\appendices

\section{Proof of Theorem \ref{thm:limit_formula_entropy}}\label{app:proof:thm:limit_formula_entropy}

We prove first an auxiliary result.

\begin{lemma}
The Borel $\sigma$-algebra $\salg B(\Rr^d)$ equals $$\salg F := \sigma(\set{C_{2^n,\V j}}{n\in \Nn, \, \V  j \in \Zz^d}).$$ 
Similarly, 
\begin{equation}\label{eq:seprability_BE}
\salg B(E) = \sigma(\set{C_{2^n,\V j}\cap E}{n\in \Nn, \, \V  j \in \Zz^d}):=\salg F'.
\end{equation}
\end{lemma}
\begin{IEEEproof}
Since each $C_{2^n,\V j}$ clearly belongs to $\mathcal \salg B(\Rr)^{\otimes d}$, which equals $\salg B(\Rr^d)$ \cite[Se.8.5]{Williams1991}, we have the inclusion $\salg F\subset \salg B(\Rr^d)$; similarly, $C_{2^n,\V j}\cap E$ belongs to $\salg B(E)$ so $\salg F'\subset \salg B(E)$. 
 
 To prove the opposite inclusion, we remark first that all norms on $\Rr^d$ are equivalent and therefore that $\salg B(\Rr^d)= \sigma(\tau)$ for the unique topology $\tau$ generated the norms. We may choose the norm 
\begin{equation}
\norm{(x_1,...,x_d)}_{\infty} = \max_{1\leq i \leq d}|x_i|,
\end{equation} 
so that the open balls are cubes 
\begin{equation}\label{eq:cube}
C= (a_1,b_1)\times \cdots \times (a_d,b_d)
\end{equation}
Remark that an open interval  $(a,b)\subset \Rr$ can be expressed as an increasing limit of sets
\begin{equation}
(a,b) = \lim_n \bigcup_{j\in \Zz\,:\, a < j2^{-n} \leq b-2^{-n}} \left[\frac{j}{2^n}, \frac{j+1}{2^n}  \right),
\end{equation}
 because for each $n$ the distance between $a$ (resp. $b$) and the leftmost (resp. rightmost) point of $U_{a,b,n}=\bigcup_{j\in \Zz\,:\, a < j2^{-n} \leq b-2^{-n}} \left[\frac{j}{2^n}, \frac{j+1}{2^n}  \right)$ is at most $2^{-n}$. Reordering products and unions in $U_{a_1,b_1,n}\times \cdots \times U_{a_d,b_d,n}$ , we conclude that the open cube $C$ in \eqref{eq:cube} is an increasing limit of a union of semi-open cubes of the form $C_{2^n,\V j}$, hence an element of $\salg F$. Because the open cubes  generate the Borel $\sigma$-algebra, 
 it follows that $\salg B(\Rr^d)\subset \salg F$. The same argument shows that $C\cap E$ is an increasing limit of union of sets of the form $C_{2^n,\V j}\cap E$, hence $\salg B(E) \subset \salg F'$.
\end{IEEEproof}

The proof of Theorem \ref{thm:limit_formula_entropy}  uses martingale techniques, relying on the fact that the partition $\{C_{2^{n+1},\V j}\}_{\V j\in \Zz^d}$ refines $\{C_{2^{n},\V j}\}_{\V j\in \Zz^d}$. 

\begin{IEEEproof}[Proof of Theorem \ref{thm:limit_formula_entropy}]

Part \ref{thm:limit-formula:part1}:  Let us introduce, for each $n\in \Nn^*$, the partition 
\begin{equation}
\salg P_n = \bigset{ C_{2^n,\V j} }{\V j\in \Zz^d},
\end{equation}
and the $\sigma$-algebra $\salg F_n = \sigma(\salg P_n)$. By the previous lemma,  $\salg F_\infty := \sigma(\bigcup_n \salg F_n) = \salg F = \salg B(E)$; in other words, $\salg B(E)$ is a separable $\sigma$-algebra.

We can thus reproduce the argument in \cite[Sec. V.56]{Dellacherie1982B}. We suppose first that $\mu$ is a probability measure.  The process $(X_n)_{n\geq 1}$ given by 
\begin{equation}
X^{\rho,\mu}_n\equiv X_n := \sum_{\substack{C\in \salg P_n\\\mu(C)>0}} \frac{\rho(C)}{\mu(C)} \cf_C
\end{equation}
is obviously adapted with respect to $\mathcal F_n$ and a positive martingale with respect to the law $\mu$ because
\begin{align}
\int_{C_{2^n,\V j}}   X_{n+1} \d\mu &=\sum_{\substack{C_{2^{n+1},\V k} \subset C_{2^n,\V j}\\\mu(C_{2^{n+1},\V k})>0}} \int_{C_{2^{n+1},\V k}} \frac{\rho(C_{2^{n+1},\V k})}{\mu(C_{2^{n+1},\V k})}\d \mu \\
&= \sum_{\substack{C_{2^{n+1},\V k} \subset C_{2^n,\V j}\\\mu(C_{2^{n+1},\V k})>0}} \rho(C_{2^{n+1},\V k}) \\
&= \rho(C_{2^{n},\V j}) = \int_{C_{2^n,\V j}} X_n\d \mu.
\end{align}
The first equality exploits that $\{C_{2^{n+1},\V j}\}_{\V j\in \Zz^d}$ refines $\{C_{2^{n},\V j}\}_{\V j\in \Zz^d}$; the  last two use absolute continuity of $\rho$ w.r.t. $\mu$. 

By the Martingale Convergence Theorem, $(X_n)$ converges $\mu$-a.s. to an $\salg F$-measurable, $\mu$-integrable random variable $X_\infty$. Since $\rho \ll \mu$, the martingale $(X_n)$ is uniformly integrable (see \cite[Sec. V.56]{Dellacherie1982B}), hence $X_n\to X_\infty$ in $L^1$ and $X_n=\mathbb E(X_\infty|\salg F_n)$. Since the relation $\rho(A) = \int_A X_\infty \d\mu$ holds for every $A\in \bigcup_n \salg F_n$, it also holds for $A\in \salg F_\infty = \salg B(E)$, thus $X_\infty = \diff\rho\mu$.

If $\mu$ is only a finite measure, we replace it first by $\tilde \mu = \mu(E)^{-1} \mu$. Then $\d\tilde \mu/\d\mu = \mu(E)^{-1}$. By the previous result, we have the almost-sure point-wise equality
\begin{equation}
\diff{\rho}{\tilde\mu} = \lim_n X_n^{\rho,\tilde\mu} = \mu(E) \lim_n X_n^{\rho,\mu},
\end{equation}
hence
\begin{equation}
\diff{\rho}{\mu} = \diff{\rho}{\tilde\mu}\diff{\tilde\mu}{\mu} = \lim_n X_n^{\rho,\mu}.
\end{equation}

More generally, if $\mu$ is $\sigma$-finite, we start partitioning the space $E$ into countably many sets $S_i$ of finite $\mu$-measure and then apply the argument above to each part, finding $\diff\rho\mu|_{S_i}$. 

Part \ref{thm:limit-formula:part2}: Remark first that if $|f|\leq B$, then the function $I(x)=x \ln (1/x) \cf_{\{x>0\}}$ is upper-bounded by $\max(\e^{-1}, B |\ln B|)$, which is integrable on $(E,\salg B(E),\mu)$. We have
\begin{align*}
\ent\mu(\rho) & =\int_E I(\diff\rho\mu)\d\mu =
 \int_{E} I(\lim_n X_{n}^{\rho,\mu}) \d\mu \\
&  = \int_{E} \lim_n I( X_{n}^{\rho,\mu}) 
 = \lim_n \int_{E}  I( X_{n}^{\rho,\mu})
\end{align*}
where the first equality holds by definition, the second corresponds to the almost sure convergence deduced in part  \ref{thm:limit-formula:part1}, the third is by continuity of $I$, and the fourth is implied by Lebesgue's dominated convergence theorem.

To finish the proof, simply remark that
\begin{align*}
\int_E I(X_n^{\rho,\mu} )\d\mu & = \sum_{\substack{\V j\in \Zz^d\\\rho(C_{2^n,\V j})>0}} \int_{C_{2^n,\V j}} \frac{\rho(C_{2^n,\V j})}{\mu(C_{2^n,\V j})} \ln \frac{\mu(C_{2^n,\V j})}{\rho(C_{2^n,\V j})} \d\mu \\
&= \ent\#(\V X_n) + \sum_{\substack{\V j\in \Zz^d\\\rho(C_{2^n\V j})>0}} p_{2^n,\V j} \ln \mu(C_{2^n,\V j}).
\end{align*}
\end{IEEEproof}

\section*{Acknowledgements}
I want to thank  Prof. Daniel Bennequin, Prof. Stephan Luckhaus, and Prof. Matilde Marcolli for some illuminating conversations on the topics covered here. The usual disclaimer is pertinent: I am solely responsible for any mistake that this article might contain.

\bibliographystyle{IEEEtran}
\bibliography{IEEEabrv,bibtex}

\end{document}